\documentstyle[12pt, epsf]{article}
\newcommand\ZZZ{{\hbox{ Z\kern-1.6mm Z}}}
\newcommand{\beq}{\begin{equation}}
\newcommand{\eeq}{\end{equation}}
\newcommand{\bea}{\begin{eqnarray}}
\newcommand{\eea}{\end{eqnarray}}
\newcommand{\ra}{{\rm }\rangle}
\newcommand{\la}{\langle}
\newcommand{\lt}{\left}
\newcommand{\rt}{\right}
\newcommand{\Iop}{\relax{\rm I\kern-.18em I}}
\newcommand{\one}{{\hbox{ 1\kern-1.2mm l}}}

\newcommand{\g}{\gamma}
\newcommand{\gb}{\bar \gamma}
\newcommand{\sig}{\sigma}

\newcommand{\del}{\partial}

\newcommand{\St}{\tilde S}

\newcommand{\CS}{{\cal S}}

\newcommand{\D}{\Delta}

\newcommand{\M}{{\cal M}}

\newcommand{\lam}{\lambda}

\newcommand{\K}{{\cal K}}

\newcommand{\th}{\theta}

\newcommand{\sectiono}[1]{\section{#1}\setcounter{equation}{0}}
\newcommand{\subsectiono}[1]{\subsection{#1}}

\begin{document}
{}~
{}~
\hfill\vbox{\hbox{DAMTP-2006-111}
\hbox{hep-th/0611138}}\break

\vskip .6cm

\centerline{\Large \bf Tachyon Condensation and Non-BPS D-branes}
\centerline{\Large \bf in a Ramond-Ramond Plane Wave Background}

\medskip

\vspace*{4.0ex}

\centerline{\large \rm Partha Mukhopadhyay }

\vspace*{4.0ex}

\centerline{\large \it Department of Applied Mathematics and
Theoretical Physics}
\centerline{\large \it University of Cambridge}
\centerline{\large \it Wilberforce Road, Cambridge CB3 0WA, UK}

\medskip

\centerline{E-mail: P.Mukhopadhyay@damtp.cam.ac.uk}

\vspace*{5.0ex}

\centerline{\bf Abstract} \bigskip

We consider string theory in maximally supersymmetric type IIB plane
wave background with constant five form Ramond-Ramond flux (RR plane
wave). It is argued that there exists a universal sector of string
configurations independent of the null coordinate $x^-$ such that
the space-time action evaluated at such a configuration is same in
the RR plane wave and flat background. By naturally assuming its
validity for the open strings we further argue that the D-branes
extending along $x^{\pm}$ are universal in these two backgrounds.
Moreover, a universal D-brane which is BPS in flat space must be
tachyon-free in RR plane wave and a non-BPS D-brane should have a
real tachyon whose potential is universal. Given the above
observation we then proceed to describe open string theories for the
non-BPS D-branes in RR plane wave. It is suggested that the
light-cone Green-Schwarz fermions on the world-sheet satisfy certain
bi-local boundary condition similar to that corresponding to flat
space. We attempt a canonical quantisation with this boundary
condition which gives rise to an open string spectrum similar to
that in flat space - containing an R and NS sectors of states. In
this process we encounter certain subtleties involving the R sector
zero modes. Although the derivation of the R sector zero mode
spectrum using the open string theory has not been completely
settled, we suggest definite answer by requiring consistency with
the relevant space-time interpretation. We finally generalise the
above basic features of universality to all the exact pp-wave
backgrounds with the same dilaton profile in a given string theory.

\newpage

\tableofcontents

\baselineskip=18pt

\sectiono{Introduction and summary}
\label{s:intro}

Open string tachyon condensation\footnote{By tachyon condensation we
shall always refer to {\it open} string tachyon condensation in this
paper.} is an interesting and important subject of study in string
theory. Its basic content is described by the well known conjectures
due to Sen\footnote{The reader is referred to the latest review
\cite{sen04} for the details of the developments in this subject and
complete list of original references.}. Even though the associated
physical picture is quite simple its complete string theoretic
understanding is still lacking.
The main obstacle is the mathematical complications of string theory
which is also responsible for the fact that it has mainly been studied
in flat space. It is therefore potentially interesting to study tachyon
condensation in other geometric backgrounds in string
theory\footnote{See \cite{lit} for relevant studies in this direction.}.
In this regard one recalls that Sen argued \cite{universal} the universality
of tachyon potential in a certain class of superstring backgrounds
which admit a symmetry under $(-1)^{F_L}$, where $F_L$ is the number
of space-time fermions that are left-moving on the world-sheet.
In Sen's argument this symmetry was needed to define a coincident
brane-anti-brane system. But there are other interesting backgrounds,
namely the Ramond-Ramond (RR) backgrounds which do not admit such a
symmetry. These backgrounds are important in string theory
because of their role in the formulation of AdS/CFT correspondence
\cite{ads/cft} and the fact that sting theory is poorly understood in
such cases.

Among the non-trivial gravitational backgrounds plane waves are special
as one gets more technical control on the quantisation problem in
light-cone gauge. In this paper we focus on the particular example of
type IIB maximally supersymmetric plane wave background with
constant RR five form flux (RR plane wave) \cite{blau01}. This is the
simplest non-trivial example \cite{metsaev01, metsaev02, bmn} that we
would like to consider for our purpose. In light-cone Green-Schwarz (GS)
formalism the
world-sheet theory \cite{metsaev01, metsaev02} is simply a mass
deformation of that corresponding to the flat background \cite{gsw}.
The mass parameter is
given by a product of the strength of the background flux and the
momentum $p_-$ conjugate to the null coordinate $x^-$ associated to
the covariantly constant null Killing symmetry of the background. We
point out that this indicates the following is true:
\bea
&& \hbox{\it There exists a universal sector of string configurations
  independent of $x^-$ such that} \cr
&& \hbox{\it the space-time action evaluated at such a configuration
  is same in RR plane wave}\cr
&& \hbox{\it and flat background.}
\label{statement}
\eea
We shall present argument in favour of this conclusion in
sec.\ref{s:us}. Although
our argument does not precisely specify what this universal
sector is\footnote{This sector will be studied in more detail in
  \cite{us}. See sec.\ref{s:generalization} for further comments on
  it.}, we argue that any scalar field configuration independent of $x^-$
should be included in it. By naturally assuming the validity of the
above statement for the open string theories we point out that it has
the following consequence for the D-branes:
\bea
&& \hbox{\it In type IIB string theory, for every D-brane in flat
  background extending along $x^{\pm}$ } \cr
&&\hbox{\it directions there exists a corresponding D-brane in RR plane
  wave background.}
\label{statement-D-brane}
\eea
This means that the D-branes extending along $x^{\pm}$ are {\it universal}.
There are two categories of such D-branes, namely {\it even} and {\it
  odd}, depending on the dimensionality of the worldvolume. As is well
known, the even D-branes are half-BPS in flat space \cite{bps}
while the odd ones break all the supersymmetries
\cite{nbps}. Although the detailed descriptions of
such universal D-branes may be different in the two backgrounds,
existence of the universal sector of string configurations predicts
that the following must be true:
\bea
&& \hbox{\it A universal D-brane in RR plane wave is tachyon free if
  it is even and contains a} \cr
&& \hbox{\it real tachyon if it is odd. Moreover the tachyon potential
  is universal.}
\label{statement-tachyon}
\eea
D-branes in RR plane wave have already been analysed by many authors
from various points of view \cite{dabholkar02, bergman02,
  skenderis02, ws, le}. According to the above classification these
studies include only the even D-branes. Indeed all such D-branes
have been found to be tachyon-free independently of their
supersymmetries. One of the main goals of this paper is to develop
the light-cone open string theories for the odd (non-BPS) D-branes.
In flat space there are D-branes for which a space-like combination
of the light-cone directions satisfies Dirichlet boundary condition
while the orthogonal time-like combination satisfies Neumann
boundary condition. Such D-branes can not be analysed in light-cone
gauge and we do not discuss the analogues of such D-branes, if they
exist, in RR plane wave.

Let us now discuss the non-BPS D-branes. In flat space the open string theories
for such D-branes are not as straightforward as those corresponding to the even ones.
In a manifestly supersymmetric formalism the complication lies in writing
down the boundary condition for the space-time spinors. This problem
was studied in light-cone gauge from both open and closed string
points of view in \cite{nbps1, bs, nbps2}. In particular, it was found
in \cite{nbps2} that the $SO(8)$ covariant open string boundary
condition is actually bi-local which relates quadratic combinations of
left and right moving GS fermions. Although the obvious
question of how to deal with such a boundary condition was not
completely settled, some progress was made. It was shown that this boundary
condition leads to two sectors of open strings (see also \cite{yoneya99})
namely, Ramond (R) and
Neveu-Schwarz (NS) sectors. The resultant open string spectrum
successfully passed through the world-sheet open-closed duality
check\footnote{Moreover, a proposal for how to compute disk amplitudes
with all possible bulk insertions, but without any boundary insertion,
was suggested \cite{nbps2} which produced correct results for the
closed string one point functions. This boundary condition was
generalised to its $SO(9,1)$ covariant form and applied to pure spinor
formalism \cite{berkovits} in \cite{ps1, ps2}.}.

In this paper we show that all the valid open string boundary
conditions for the kappa symmetry fixed type IIB GS
superstrings in flat background are also valid in RR plane wave.
This justifies us to use the similar bi-local boundary
condition as discussed above to define the non-BPS D-branes in RR plane
wave. It is shown that, just like in flat space \cite{nbps2}, there
exist an R and NS sectors of open strings. However,
there are certain additional subtleties involved in the present case.
Usually a boundary condition enables us to solve for a set
of degrees of freedom in terms of the rest. In absence of a linear
boundary condition it is not {\it a priori} clear if the world-sheet
lagrangian can be written down entirely in terms of the independent
modes. We show that this can indeed be done almost for all the degrees
of freedom except for the R sector zero modes - a complication which
was absent in flat space.

We shall now briefly discuss this zero mode problem. Because of the
mass term, the equations of motion dictate that the zero modes be
either $\tau$ (world-sheet time) or $\sig$ (world-sheet space)
dependent. It turns out that its only the $\tau$ dependent zero
modes $S^{1a}_0(\tau)$ and $S^{2a}_0(\tau)$ for which one can
possibly make sense of the proposed bi-local boundary condition. The
problem is that while the two zero modes are related quadratically,
the mass term in the lagrangian is linear in both and therefore can
not be expressed in terms of any of the two. However, since this
term does not contain any $\tau$ derivative, construction of the
hamiltonian still goes through. Finally one arrives at a particular
algebraic structure involving the constant zero modes $S^a_0$ and
$\St^a_0$ which appear in the classical solutions for the $\tau$
dependent zero modes. This algebraic structure turns out to be quite
unusual as $S^a_0$ and $\St^a_0$ are not linearly related. Although
we are unable to settle the issue at this point, we argue that
consistency with the relevant low-energy effective action requires
us to have the same zero-mode spectrum as that in flat space.

In the NS sector the zero point energy does not cancel between the
bosons and fermions because of broken supersymmetry. In flat space
the standard zeta function regularisation gives the correct answer
which produces a consistent spectrum for an interacting covariant
field theory in space-time. The usual approach in the literature in
the context of RR plane wave has been to consider the relevant
Casimir energy which can be computed by using a generalised zeta
function regularisation. But in this case we do not have an
independent physical justification for this result. We remain
inconclusive to this issue and leave its resolution as a future
work.

The rest of the paper is organised as follows: We present our argument
for the existence of the universal sector of string configurations in
sec.\ref{s:us}. The relevant consequences for the D-branes are
discussed in sec.\ref{s:tach}. The open string theory for the
GS fermions on a non-BPS D-brane in RR plane wave is
discussed in sec.\ref{s:open}. Finally we comment on certain
generalisations in sec.\ref{s:generalization}. Several
appendices contain necessary technical details.

\sectiono{A universal sector of string configurations}
\label{s:us}

Here we shall argue in favour of the statement made in
(\ref{statement}). The type IIB RR plane wave background is given by,
\bea
ds^2 &=& G_{\mu \nu} dx^{\mu} dx^{\nu} ~, \cr
&=& 2 dx^+ dx^- - f^2 x^Ix^I dx^+ dx^+ + dx^I dx^I ~, \quad I =
1,\cdots ,8~,\cr
F_{+1234}&=&F_{+5678} \propto f ~,
\label{RRpp-backgrd}
\eea
where $F$ is the RR five form field strength. Although $f$ can be
changed by properly scaling the light-cone coordinates the limit $f
\to 0$ is special which corresponds to the flat background.
String theory in this background has been solved in the standard
light-cone gauge in \cite{metsaev01, metsaev02} which we refer the
reader to for details. We shall follow the notations of
\cite{metsaev02} for our argument in this section and use them without
any introduction.

The world-sheet theory in light-cone gauge is a mass deformation
of that in flat background where the mass is given by\footnote{We use
  the notation $p_-$ for the light-cone momentum which is same as
  $p^+$ used in \cite{metsaev02}.},
\bea
m=2 \pi \alpha' p_- f~.
\label{m}
\eea
Although the bosonic sector of the theory maintains the complete
transverse symmetry under $SO(8)$, the fermionic mass term breaks it
to $SO(4) \times SO(4)$ as it is evident from the appearance of the
matrix $\Pi$ \cite{metsaev02} which is the product of four gamma
matrices along the
directions: $1, \cdots, 4$. This breaking of the global symmetry is
expected due to the presence of the background flux.

Notice that the light-cone world-sheet theory is only sensitive to $m$
which contains a product of $p_-$ and $f$. This gives an indication
that the $f\to 0$ limit is mimicked in the sector where the space-time
fields are independent of the null coordinate $x^-$ and therefore the
background appears to be flat. To see this in a little more detail let
us consider the space-time action in the background
(\ref{RRpp-backgrd}) written in terms of the Fourier transformed
fields. A generic term in this action can be schematically written as,
\bea
T_{pw} = \int \delta^{10}(\sum_i p^i)
{\cal A}(f, p_-^i, p_+^i, p_I^i) \prod_i \chi_i(p_-^i,p_+^i,p_I^i)~,
\label{Tpw}
\eea
where the integration is over all the momenta $p^i$. $\chi_i$'s are
the Fourier transformed fields. They carry space-time indices and can
either be bosonic or fermionic. In general ${\cal A}$ is a function of
all the momenta and the background parameter $f$. There are space-time
index contractions among $\chi_i$'s and ${\cal A}$ which are
suppressed in this schematic expression. If ${\cal A}$ survives the
$f\to 0$ limit then,
\bea
T_{flat} = \int \delta^{10}(\sum_i p^i) {\cal A}(0, p_-^i, p_+^i,
p_I^i) \prod_i \chi_i(p_-^i,p_+^i,p_I^i)~,
\label{Tflat}
\eea
gives a term in the space-time action in flat background. If ${\cal
  A}$ does not survive the $f\to 0$ limit then it corresponds to
certain interaction that is absent in flat space. If this limit has to
be automatically mimicked in the sector where all the $p_-^i$'s are
set to zero then we should have,
\bea
{\cal A}(f, p_-^i=0, p_+^i, p_I^i) = {\cal A}(0, p_-^i=0,p_+^i, p_I^i)~.
\label{A-cond}
\eea
Since we are assuming a smooth flat space limit for ${\cal A}$, it can
be Taylor expanded and only positive powers of $f$ appear.
Also usually ${\cal A}$ follows some power law behaviour in terms of
the momenta. With these assumptions the above equation implies,
\bea
{\cal A}(f, p_-^i, p_+^i, p_I^i)={\cal A}'(m^i,p_-^i, p_+^i,p_I^i)~,
\label{A-A'}
\eea
where ${\cal A}'$ is another function and $m^i=2 \pi \alpha'p^i_- f$.
This means
that $f$ always appears with a factor of $p^i_-$. ${\cal A}$ has to be
computed using the world-sheet theory and we have already seen that
this condition is indeed maintained in the light-cone gauge. Notice
that the above argument does not precisely specify the universal
sector that was referred to in (\ref{statement}). Although the
required condition (\ref{A-cond}) or (\ref{A-A'}) is satisfied in
light-cone gauge, it is not clear what should be the description of
the universal sector outside the light-cone gauge.
Work is in progress in \cite{us} to answer such
questions. Nevertheless a scalar field is certainly included in this
sector and this is what we need for our argument on tachyon
condensation in the next section.

We shall now discuss the symmetry algebra which is not same as that of
the flat background. The kinematical and dynamical generators in the
RR plane wave are \cite{metsaev02}:
$P^+$, $P^I$, $J^{+I}$, $J^{ij} ~(i,j = 1,\cdots ,4)$,
$J^{i'j'} ~(i',j'=5,\cdots ,8)$, $Q^{+A} ~(A=1,2)$ and $P^-$, $Q^{-A}$
respectively. In flat space the list includes certain extra
kinematical generators, namely: $J^{+-}$, $J^{-I}$ and
$J^{ii^{\prime}}$, which are missing in RR plane wave. Also the
matching generators have different expressions in the two
backgrounds. We therefore need to
\begin{enumerate}
\item
account for the missing generators,
\item
show that in expressions for the matching generators and algebra in RR
plane wave $f$ appears only as $m$.\footnote{One should also expect
  that the $\Pi$ dependence should go away in the limit $p_-\to
  0$. This is automatically satisfied as $f \to 0$ is the flat space
  limit and therefore $\Pi$ should always come with factors of $f$.}
\end{enumerate}
To account for the missing generators $J^{ii^{\prime}}$we notice that
the matching generators $J^{ij}$ and $J^{i^{\prime}j^{\prime}}$ do not
depend on $f$ (hence $\Pi$).
Their construction can, in fact, be algebraically continued to
incorporate $J^{ii^{\prime}}$. The other missing generators $J^{+-}$
and $J^{-I}$ do not preserve the light-cone gauge and our analysis
does not go beyond this gauge\footnote{I thank Peter van Nieuwenhuizen
  for asking pertinent questions on this.}. To analyse this we need a
covariant setup with a precise knowledge of the universal sector in
(\ref{statement}). But one can make the following comments for
$J^{+-}$. This generates equal and opposite scaling of
the coordinates $x^+$ and $x^-$. Certainly the background
(\ref{RRpp-backgrd}) is not invariant under such a transformation as
it rescales the flux-strength $f$. But since $f \to 0$ limit is
mimicked in the universal sector, such a transformation should appear
as a symmetry in this sector.

To address the second point above we notice that the dynamical
generators do depend on $f$ and indeed the dependence is only through
$m$. This, however, is not true of symmetry algebra - $f$ explicitly
appears in commutation relations. Schematically, $f$ appears in the
following way in such relations \cite{metsaev02},
\bea &&
\lt[P^-, P^I\rt] \sim f^2 J^{+I}~,\lt[P^-,Q^+ \rt] \sim f Q^+
~,\lt[P^I,Q^-\rt] \sim f Q^+~, \cr
&& \lt\{Q^+,Q^-\rt\} \sim f
J^{+I}~,\lt\{Q^-,Q^-\rt\} \sim f J^{IJ}~.
\label{alg}
\eea
But generators have additional $p_-$ dependence:
\bea
&& P^+=p_-~, P^- \sim {1\over p_-}~, J^{+I} \sim
p_-~, \cr
&& Q^+ \sim \sqrt{p_-} ~, Q^- \sim {1\over
\sqrt{p_-}}~.
\label{p+dep}
\eea
These $p_-$ dependence are such that they can be used in
eqs.(\ref{alg}) to absorb all the $f$ dependence and write the
relevant commutation relations in terms of $m$.

\sectiono{Tachyon condensation and D-brane descent relation in RR plane wave}
\label{s:tach}

We shall now consider (\ref{statement}) in the context of D-branes and
discuss its possible consequences for the tachyon condensation in RR
plane wave. In particular, we shall argue below that the statement
(\ref{statement-D-brane}) is expected to be true.

To demonstrate the idea with the simplest example let us consider a
non-BPS D-brane, D$p$ with $p$ even, in type IIB string theory in flat
background. Furthermore we orient it to extend along both the
light-cone directions so that there are $(p-1)$ light-cone transverse
directions (collectively denoted as $\vec x$) along the
worldvolume. The worldvolume theory contains a real tachyon
$T(x^+,x^-,\vec x)$. This has a potential $V(T)$ so that the
worldvolume action evaluated for a field configuration where the
tachyon is set to a constant $T$ and all the rest of the fields are
set to zero is given by,
\bea
S_{flat}[T] = - {\cal V}^{(p+1)} V(T)~,
\eea
where ${\cal V}^{(p+1)}=\int dx^+ dx^- d\vec x$ is the (infinite)
volume of the worldvolume. $V(T)$ is an even function of $T$ such
that: (1) it has a maximum at $T=0$ where the hight of the potential
is same as the tension of the D-brane, (2) it has non-perturbative
minima at $|T|=T_0$ which correspond to the closed string vacuum as
conjectured by Sen.  It is natural to expect that $S_{flat}$ can be
obtained from a covariant action $S_{cov}$ relevant to type IIB string
theory by expanding it around the flat background. Given $S_{cov}$ it
should also be possible to expand it around the RR plane wave
background (\ref{RRpp-backgrd}) to obtain $S_{pw}$. As we have argued
in the previous section that tachyon, being a scalar field, is
included in the universal sector in (\ref{statement}). Therefore we
must have,
\bea
S_{pw} [T] = - {\cal V}^{(p+1)} V(T)~.
\eea
Now recall \cite{sen04} that a suitable BPS D$(p-1)$-brane in flat
space can be viewed as a tachyonic kink solution $T(y)$ in the
worldvolume theory described by $S_{flat}$.
\bea
T(y) = \lt \{
\begin{array}{ll}
\pm T_0 ~, & \hbox{as } y\to \pm \infty ~, \\
0 ~, & \hbox{at } y=0 ~,
\end{array} \rt.
\label{kink}
\eea
where $y$ is one of the light-cone transverse directions along the
worldvolume of the parent D-brane that is transverse to the lower
dimensional BPS D-brane. Notice that any solution of the above kind is
necessarily a solution of the theory described by the action $S_{pw}$
as (\ref{statement}) applies to such configurations. Therefore one
would expect this solution to correspond to an even D-brane in RR
plane wave.

The basic argument used above also applies to the worldvolume theories
on brane-anti-brane systems. Any D-brane in flat space that extends
along $x^-$ can be realised as a tachyonic solution that is
independent of $x^-$ in the worldvolume theory of sufficiently large
number of brane-anti-brane systems or non-BPS D-branes (see
\cite{sen04} for the detailed discussion and references therein.).
According  to the above argument any such solution
belongs to the universal sector of some worldvolume theory and
therefore the corresponding D-brane is expected to exist in RR plane
wave. This
establishes a correspondence between the relevant D-branes in flat and
the RR plane wave background as mentioned in
(\ref{statement-D-brane}). The only assumption in this argument is the
existence of a covariant completion of any worldvolume theory that one
obtains from open strings in flat space.

Although the above discussion implies that any universal D-brane is
represented by the same worldvolume solution in flat background and RR
plane wave, their detailed properties are in general different. For
example, it is not guaranteed that a given universal D-brane preserves
the same amount of supersymmetry in the two backgrounds. However,
there are certain properties, such as those mentioned in
(\ref{statement-tachyon}), are guaranteed to be restricted by the
existence of the universal sector. Referring the reader to
(\ref{statement-tachyon}), we comment that it is indeed true that all
the even D-branes that have so far been discussed in the literature
\cite{dabholkar02, bergman02, skenderis02, ws, le} are found to be
tachyon-free. Given this discussion the next obvious question is:
what is the open string theory for a non-BPS D-brane in RR plane wave
that is predicted by universality? We suggest an answer to this question
in the next section.

\sectiono{Non-BPS D-branes in RR Plane Wave}
\label{s:open}

Here we shall discuss the open string theory for non-BPS D-branes in
type IIB RR plane wave background in light-cone gauge. We have shown
in appendix \ref{a:openbc} that all the known open string boundary
conditions in kappa gauge fixed GS action in flat
background are also valid boundary conditions in RR plane
wave\footnote{Following \cite{ps1} we have also written down the
  $SO(9,1)$ covariant boundary condition for the space-time spinors
  for a non-BPS D-brane in flat space using the covariant
  GS action.}. This justifies considering the same boundary
conditions for the universal D-branes in both flat space and RR plane
wave. Therefore in the bosonic sector
we have the standard Neumann/Dirichlet boundary conditions - their
analysis is quite standard and we shall not discuss them in detail. As
in flat space the actual subtlety is involved in the space-time
fermionic part \cite{nbps1, bs, nbps2} and therefore we shall focus
our attention only to this sector in the following discussion.

The relevant world-sheet lagrangian is of the following form,
\bea
L = \beta \int_0^{\pi} d \sigma \lt[ \lt(S^1\del_+S^1\rt) +
  \lt(S^2\del_-S^2 \rt) - 2m \lt(S^1\Pi S^2\rt)\rt]~,
\label{LF}
\eea
where $\beta$ is a constant and following the gamma matrix convention
of \cite{ps1} we have,
\bea
\Pi_{ab} = \sig^{1\cdots 4}_{ab} = \sig^{5\cdots 8}_{ab}~.
\label{Pi}
\eea
The equations of motion read,
\bea
\del_+ S^1-m \Pi S^2 = 0 ~, \quad \del_- S^2 + m \Pi S^1 =0~,
\label{eom}
\eea
The boundary condition, as discussed in appendix \ref{a:openbc},
is given by \cite{nbps2},
\bea
S^{1a}(\tau, \sigma) S^{1b}(\tau', \sigma) = \M^{ab}_{cd} S^{2c}(\tau,
\sigma) S^{2d}(\tau', \sigma)~, \quad \hbox{at } \sigma = 0,\pi ~,
\label{nBPSbc-S1S1}
\eea
where the expression for ${\cal M}^{ab}_{cd}$ can be found in
eq.(\ref{calM-lc}). Since the fields are massive the equations of
motion (\ref{eom}) mix $S^1$ and $S^2$. Indeed using these equations
of motion one can write down the following set of all possible
boundary conditions,
\bea
S^{Aa}(\tau, \sigma) S^{Bb}(\tau',\sigma) = \M^{ABab}_{CDcd}
S^{Cc}(\tau ,\sigma) S^{Dd}(\tau',\sigma)~, \quad \hbox{ at } \sigma =0,\pi~,
\label{nBPSbc-SASB}
\eea
where the non-zero entries for $\M^{ABab}_{CDcd}$ are given by:
\bea
\M^{11ab}_{22cd} = \M^{22ab}_{11cd} = \M^{ab}_{cd} ~, \quad
\M^{12ab}_{21cd} = \M^{21ab}_{12cd} = -\lt[\M\lt(\Pi\over
  \Pi\rt)\rt]^{ab}_{cd}~,
\label{MABabCDcd}
\eea
where we have used the notation:
$\displaystyle{\lt[\M\lt(\Pi \over \Pi\rt)\rt]^{ab}_{cd} =
  \M^{aa'}_{cc'}\Pi_{a'b} \Pi_{c'd}}$. Since $\M^{ab}_{cd}$ has four
indices $\Pi_{ab}$ can be contracted in four different ways. The
position of $\Pi$ in the above notation indicates which index of $\M$
it is attached to\footnote{For example there is only one $\Pi$
  attached to $\M$ in
$\displaystyle{\lt[\M\lt(\Pi \rt)\rt]^{ab}_{cd}}$ and three in
$\displaystyle{\lt[\lt(1\over \Pi \rt)\M\lt(\Pi \over \Pi
    \rt)\rt]^{ab}_{cd}}$.}. The fields on the doubled surface
(cylinder) can be defined similarly as in (\ref{doublingTh}),
\bea
\CS^{Aa}(\tau, \sigma) \CS^{Bb}(\tau', \sigma') =
\lt \{ \begin{array}{ll}
S^{Aa}(\tau,\sigma) S^{Bb}(\tau',\sigma') & 0\leq \sigma, \sigma' \leq
\pi~, \cr
\M^{ABab}_{CDcd} S^{Cc}(\tau,2\pi-\sigma) S^{Dd}(\tau',2\pi-\sigma') &
\pi \leq \sigma, \sigma' \leq 2\pi~.
\end{array} \rt.
\label{doublingS}
\eea
Using (\ref{doublingS}), (\ref{MABabCDcd}) and (\ref{nBPSbc-SASB}) one
can show,
\bea
\CS^{Aa}(\tau, 2\pi) \CS^{Bb}(\tau',2\pi) = \CS^{Aa}(\tau,0)\CS^{Bb}(\tau',0)~,
\eea
which implies,
\bea
\CS^{Aa}(\tau, 2\pi) = \pm \CS^{Aa}(\tau, 0)~.
\eea
Therefore there exist two sectors (just like in flat space
\cite{nbps2}): R and NS, corresponding to the above signs
respectively. We shall now discuss the canonical quantisation for
these two sectors separately below.

\subsectiono{R Sector}
\label{ss:R}

R sector allows zero modes which can either be $\tau$ or $\sig$
dependent. As discussed in appendix \ref{a:even} there are two classes
of even D-branes, namely class I and class II, depending on whether
the zero modes are $\tau$ or $\sig$ dependent respectively. The
resulting algebraic structures are also different. We find it hard to
use the present bi-local boundary condition to solve for dependent
modes in terms of the independent ones in case of $\sig$ dependent
zero modes. Below we shall show that with the $\tau$ dependent zero
modes a consistent canonical quantisation for the open strings can be
done with the proposed bi-local boundary condition leading to an
algebraic structure which resembles more like the class I even D-branes.

We, therefore, start out with the following mode expansion:
\bea
S^1(\tau, \sigma) &=& S_0^1(\tau) + \sum_{n\neq 0} c_n \lt\{S_n(\tau)
e^{in\sigma} + i d_n \Pi \St_n(\tau) e^{-in\sigma} \rt\}~, \cr
S^2(\tau, \sigma) &=& S^2_0(\tau) + \sum_{n \neq 0} c_n
\lt\{\St_n(\tau) e^{-in\sigma} - id_n \Pi S_n(\tau) e^{in\sigma}
\rt\}~,
\label{Rbasis}
\eea
where,
\bea
c_n = {1\over \sqrt{1+d_n^2}}~, \quad d_n ={1\over m}(w_n-n)~,
w_n=\hbox{sign}(n) \sqrt{n^2+m^2}~.
\label{cdw}
\eea
The classical equations of motion for the modes are given by,
\bea
\dot S^1_0(\tau) = m \Pi S^2_0(\tau)~, \quad \dot S^2_0(\tau) = -m
\Pi S^1_0(\tau)~, \cr
\dot S_n(\tau) = -iw_n S_n(\tau) ~, \quad \dot {\tilde S}_n(\tau) =-i
w_n \tilde S_n(\tau)~.
\label{eom-mode}
\eea
All the degrees of freedom $S^1_0(\tau)$, $S^2_0(\tau)$, $S_n(\tau)$
and $\St_n(\tau)$ are not independent because of the boundary
condition (\ref{nBPSbc-SASB}). We shall now impose these conditions
to determine the independent degrees of freedom. Naturally this task
is more cumbersome than in the usual case of an even D-brane.
Choosing $A=B=1$ and
$A=1, B=2$ in eq.(\ref{nBPSbc-SASB}) one arrives at two sets of
relations involving these degrees of freedom. It turns out that there
exists a set of
independent solutions satisfying all such relations such that all
these solutions relating various modes are independent of $m$. These
solutions are bi-local and are given by,
\bea
S^{1a}_0(\tau) S^{1b}_0(\tau^{\prime}) &=& \M^{ab}_{cd} S^{2c}_0(\tau)
S^{2d}_0(\tau^{\prime})~, \cr
S^{1a}_0(\tau) S^{2b}_0(\tau^{\prime}) &=& -\lt[ \M\lt(\Pi\over
  \Pi\rt)\rt]^{ab}_{cd} S^{2c}_0(\tau) S^{1d}_0(\tau^{\prime})~.
\label{zz}
\eea
\bea
S^{1a}_0(\tau) S^b_n(\tau^{\prime}) &=& \M^{ab}_{cd} S^{2c}_0(\tau)
\St^d_n(\tau^{\prime})~, \cr
S^{1a}_0(\tau) \St^b_n(\tau^{\prime}) &=& -\lt[ \M\lt(\Pi \over
  \Pi\rt) \rt]^{ab}_{cd} S^{2c}_0(\tau) \St^d_n(\tau^{\prime})~.
\label{zn}
\eea
\bea
S^a_n(\tau) S^b_{n^{\prime}}(\tau^{\prime}) &=& \M^{ab}_{cd}
\St^c_n(\tau) \St^d_{n^{\prime}}(\tau^{\prime})~, \cr
S^a_n(\tau) \St^b_{n^{\prime}}(\tau^{\prime}) &=& - \lt[\M\lt(\Pi
  \over \Pi\rt) \rt]^{ab}_{cd} \St^c_n(\tau)
S^d_{n^{\prime}}(\tau^{\prime})~.
\label{nnprime}
\eea
The above equations suggest that
any expression that is quadratic in the set of modes
$\{S^2(\tau), \St_n(\tau)\}$ can be expressed entirely in terms of the
set of modes $\{S^1_0(\tau), S_n(\tau)\}$ using the above
conditions. We shall see that this will almost suffice for the present
quantisation problem except for certain subtlety involving the zero
modes which will be discussed as we go on.

We shall now attempt to compute the lagrangian (\ref{LF}) in terms of
the independent modes using the relations (\ref{zz}, \ref{zn},
\ref{nnprime}). Substituting the expansion (\ref{Rbasis}) in
(\ref{LF}) one gets,
\bea
L = \beta \lt(L_0+L_1+L_2 \rt)~,
\label{LF2}
\eea
where, $L_0$, $L_1$ and $L_2$ are the parts with the degree of
non-zero modes being $0$, $1$ and $2$ respectively. We find,
\bea
{L_0 \over 2\pi} &=& {1\over 2} \lt(S^1_0(\tau)\dot S^1_0(\tau)\rt)
 + {1\over 2} \lt(S^2_0(\tau)\dot S^2_0(\tau)\rt)
-m \lt(S^1_0(\tau)\Pi S^2_0(\tau)\rt)~, \cr
&=& \lt(S^1_0(\tau)\dot S^1_0(\tau)\rt)-m \lt(S^1_0(\tau)\Pi S^2_0(\tau)\rt)~,
\label{L0}
\eea
where in the second step we have used the first equation in
(\ref{zz}). Since this condition is bi-local in nature one can take
$\tau$-derivatives at the two points independently before taking the
coincidence limit. Using the same technique with conditions (\ref{zn})
one finds,
\bea
L_1 =0~.
\label{L1}
\eea
Finally, using conditions (\ref{nnprime}) one obtains,
\bea
L_2 = 2\pi \sum_{n\neq 0} \lt[\lt(S_n(\tau)\dot{S}_{-n}(\tau)\rt) -iw_n
\lt(S_n(\tau)S_{-n}(\tau)\rt)\rt]~.
\label{L2}
\eea
Notice that the lagrangian (\ref{LF2}, \ref{L0}, \ref{L1},
\ref{L2}) is almost entirely expressible in terms of the independent
modes except for the mass term for the zero modes in (\ref{L0}) which
is linear in $S^2_0(\tau)$. We can not express
this term completely in terms of the independent modes as the conditions
(\ref{zz}) are not factorisable to linear relations. Nevertheless we
can smoothly proceed to construct the hamiltonian as this term does
not contain any $\tau$ derivative. The canonically conjugate momenta
are,
\bea
\Pi^a_0(\tau) = {\del L \over \del \dot S^{1a}_0 (\tau)} = 2\pi \beta
S^{1a}_0 (\tau)~, \quad
\Pi^a_n(\tau) = {\del L \over \del \dot S^{a}_n (\tau)} = 2\pi \beta
S^{a}_{-n} (\tau)~.
\eea
The equal time canonical anti-commutators are found to be,
\bea
\{S^{1a}_0(\tau), S^{1b}_0(\tau)\} = {i\over 2\pi \beta} \delta^{ab}~, \quad
\{S^a_n(\tau), S^b_{n'}(\tau)\} = {i\over 2\pi \beta} \delta_{n+n',0}
\delta^{ab}~.
\label{anti-comm-R}
\eea
The hamiltonian turns out to be,
\bea
H &=& H_0 -4\pi i \beta \sum_{n>0} w_n \lt(S_{-n}S_n\rt) - 8
\sum_{n>0} w_n ~, \cr
H_0 &=& 2m\pi \beta \lt(S^1_0(\tau)\Pi S^2_0(\tau)\rt)~,
\label{HF}
\eea
where the last term is the standard normal ordering constant and we
have used the $\tau$ independent oscillators for the non-zero modes
defined through the solutions of equations of motion:
\bea
S_n(\tau) &=& \exp(-iw_n\tau) S_n~, \quad \forall n \neq 0~,
\label{tau-indep-mode}
\eea
We shall now turn to the analysis of the zero modes below.

\begin{center}
{\large \bf The zero-mode problem}
\end{center}
The solutions to the zero mode equations of motion are given by,
\bea
S^1_0(\tau) = \cos(m\tau) S_0 + \sin(m\tau) \St_0~, \quad
S^2_0(\tau) = \cos(m\tau) \Pi \St_0 - \sin(m\tau) \Pi S_0 ~.
\eea
Substituting these into eqs.(\ref{zz}) we find the following relations
involving the $\tau$-independent zero-modes:
\bea
\St^a_0 \St^b_0 = \widehat{\cal M}^{ab}_{cd} S^c_0 S^d_0 ~, \quad
S^a_0 \St^b_0 = - \widehat{\cal M}^{ab}_{cd} \St^c_0 S^d_0~,
\label{StSt}
\eea
where $\widehat {\cal M}^{ab}_{cd}$ is obtained by replacing $\lam_I$
by $\hat \lam_I$ in the expression of ${\cal M}^{ab}_{cd}$ in
(\ref{calM-lc}),
\bea
\widehat {\cal M}^{ab}_{cd} =
\lt[\lt(1\over \Pi \rt)\M\lt(1\over \Pi \rt) \rt]^{ab}_{cd}
= {\cal M}^{ab}_{cd}|_{\lam_I \to \hat \lam_I}~,
\label{calMhat}
\eea
where,
\bea
\hat \lam_I = \lam_I \beta_I ~,
\label{lamhat}
\eea
and $\beta_I$ is defined in eq.(\ref{beta}). The zero-mode equal-time
anti-commutator in eqs.(\ref{anti-comm-R}) implies,
\bea
\{ S^a_0, S^b_0\} = \{ \St^a_0, \St^b_0 \} = {i\over 2\pi \beta}
\delta^{ab}~, \quad
\{S^a_0, \St^b_0 \} = \Delta^{ab}~,
\label{anti-comm-SSt}
\eea
such that,
\bea
\Delta^{ab}+ \Delta^{ba} = 0~.
\label{D-antisymm}
\eea
The first set of equations in (\ref{anti-comm-SSt}) are consistent
with the fist equation in (\ref{StSt}). Using the first equation in
(\ref{StSt}) and (\ref{D-antisymm}) one finds the zero mode
contribution to the hamiltonian to be,
\bea
H_0 = 2\pi m \beta (S_0\St_0)~.
\label{H0}
\eea
Given the algebraic structure in eqs.(\ref{StSt}, \ref{anti-comm-SSt})
and (\ref{D-antisymm}) our final goal is to construct the zero-mode
spectrum. This task is not straightforward as the relevant algebraic
structure is quite unusual. Below we shall first try to understand the
difficulty by comparing our case with that of the class I even
D-branes and then find the expected result from an argument using the
effective field theory on the worldvolume. Although it is not completely
clear to us how to get this result from the open string theory, we discuss
various arguments in appendix \ref{a:zero} for a possible resolution which
reproduces this result.

\begin{center}
{\large \it Comparison with class I even D-branes}
\end{center}
Certain relevant features of the even D-branes have been summarised in
appendix \ref{a:even}. Notice that the emergent
algebraic structure is closer to that of the class I even
D-branes than the ones belonging to class II. Equations
(\ref{anti-comm-SSt}, \ref{D-antisymm}) and (\ref{H0}) also hold in case of
class I D-branes, the only difference being that the relation between
$\St_0$ and $S_0$ is linear, unlike the ones in eqs.(\ref{StSt}).
This makes a big difference between the two cases.
First of all, for class I D-branes $H_0$ can be expressed
entirely in terms of the independent modes. Secondly, $\D^{ab}$ can be
easily computed to be a c-number which enables one to construct the
eigenstates of $H_0$. Although the spectrum of $H_0$ is independent of
$\D^{ab}$, the $H_0$ eigenstates acquire certain angular momenta
depending on $\D^{ab}$ \cite{skenderis02}. On the contrary, in our
case one can argue
that $\D^{ab}$ can not be a non-zero c-number: Since $\D^{ab}$ is
required to be anti-symmetric (see eq.(\ref{D-antisymm})) it should be
possible, if it is a c-number, to expand it in terms of the second
rank $SO(8)$ gamma matrices. The manifest covariance of the whole
construction would then require us to be able to single out two
special directions. This can only be done for the class I even
D-branes in this background because of their special $(n,n\pm 2)$-kind
of alignments (see appendix \ref{a:even}). This is not true for the
non-BPS D-branes and therefore there is no c-number choice for $\D^{ab}$
that is physically consistent in this case. This is effectively a variant of
the statement that a linear relation does not exist between $S^a_0$
and $\St^a_0$. Moreover, notice that the second relation in
(\ref{StSt}), being quadratic,
at most imposes certain constraints on $\D^{ab}$ without actually
fixing it. Therefore
the standard way of constructing the $H_0$ eigenstates does not go through.

\begin{center}
{\large \it What is expected?}
\end{center}
The fact that $\St^a_0$ could not be eliminated completely in our case
and that $\D^{ab}$ does not get fixed might lead one to think that it
is inconsistent to have these zero modes\footnote{This was initially
  thought to be the case in \cite{nBPS-RR}.}.
 This would then imply that
there exists a unique ground state instead of the usual ones that
transform as $8_{\hbox{v}}$ ($|I\ra$) and $8_{\hbox{c}}$ ($|\dot
a\ra$) under $SO(8)$.
But the existence of the universal sector of string configurations
implies that these states should be present because of the following
reason: Let us, for example, consider the vector states $|I\ra$ - they
correspond to the transverse components $A_I(x^+,x^-, \vec x)$ of the
$U(1)$ gauge field surviving in the light-cone gauge on the
worldvolume. A configuration $A_I(x^+, \vec x)$ independent of $x^-$
is inside the universal sector and therefore
should exist in RR plane wave. This is also true for the conjugate
spinor states $|\dot a\ra$. Given the above argument, the next
question will be how $H_0$ acts on these ground states.
The answer can be guessed in the following way:
As mentioned in appendix \ref{a:even}, for class I even D-branes $H_0$
breaks the degeneracy of the ground states which can be understood in the
worldvolume theory as arising from a particular Chern-Simons coupling
or its dimensional reduction. Such
couplings are not present on the worldvolume of a non-BPS D-brane and
therefore one would expect the degeneracy to be preserved in the
present case. Therefore our conclusion is that the zero-mode spectrum
is given by,
\bea
H_0 |I\ra = 0~, \quad H_0 |\dot a\ra = 0~.
\label{zero-mode-spectrum}
\eea

\subsectiono{NS Sector}
\label{ss:NS}
The analysis in the NS sector is more straightforward as there are no
zero modes. The mode expansions are,
\bea
S^1(\tau, \sig) &=& \sum_{r} c_r \lt\{ S_r(\tau) e^{ir\sig} + id_r \Pi
\St_r(\tau) e^{-ir\tau} \rt\}~, \cr
S^2(\tau,\sig) &=& \sum_{r} c_r \lt\{ \St_r(\tau) e^{-ir\sig} - id_r
\Pi S_r(\tau) e^{ir\tau} \rt\}~,
\eea
where $r$ is an half-odd-integer and $c_r$, $d_r$ and $w_r$ are given
by the same expressions as in eqs.(\ref{cdw}) with $n$ replaced by
$r$. Classical equations of motion and the quadratic relations among
the modes follow the equations for the non-zero modes given in
eqs.(\ref{eom-mode}) and (\ref{nnprime}) respectively with the similar
replacement. Performing the similar computation as in the R sector one
finally arrives at the hamiltonian:
\bea
H= -4\pi i \beta \sum_{r>0} w_r (S_{-r}S_r) - 8\sum_{r>0} w_r ~,
\eea
in terms of the $\tau$ independent modes, defined similarly as in
(\ref{tau-indep-mode}), which satisfy the following anti-commutation relation:
\bea
\{ S^a_r, S^b_s \} = {i\over 2\pi \beta} \delta_{r+s, 0} \delta^{ab}~.
\label{anti-comm-NS}
\eea

\subsectiono{Open string spectrum}
\label{ss:spectrum}

Based on the discussion in the previous subsections we are now in a
position to spell out what the open string spectrum is.
For a non-BPS D-brane sitting at the origin of the transverse directions
the light-cone hamiltonian in R and NS sectors are given by,
\bea
H^R= H_B + H^R_F ~, \quad H^{NS}=H_B + H_F^{NS}~,
\eea
respectively where the bosonic part is given by,
\bea
2 \alpha' p_-H_B = {m\over \pi} a^{I_{\parallel}\dagger}a^{I_{\parallel}} +
{m\over 2 \pi} (p-1) + \sum_{n>0} \alpha^I_{-n} \alpha^I_n ~.
\label{HB}
\eea
Recall the expression for $m$ given in eq.(\ref{m}). $I_{\parallel}$
runs over the light-cone transverse directions that are parallel to
the world-volume. The oscillators $a^{I_{\parallel}}$ are constructed
out of the bosonic zero modes and $\alpha^I_n$'s are the standard
bosonic non-zero modes. The commutation relations are:
\bea
\lt[a^{I_{\parallel}}, a^{J_{\parallel}\dagger} \rt] =
  \delta^{I_{\parallel} J_{\parallel}}~, \quad \lt[ \alpha^I_n,
    \alpha^J_{n'} \rt] = \delta^{IJ} w_n \delta_{n+n',0}~.
\eea
The second term on the right hand side of eq.(\ref{HB}) is simply the
normal ordering constant coming from the bosonic zero mode
oscillators. The ground states in the R sector are $|I\ra$
and $|\dot a\ra$ which are
annihilated by $a^{I_{\parallel}}$, $\alpha^I_n$ and $S^a_n$ for
$n>0$. The unique ground state $|0\ra$ in the NS sector is, as usual,
defined to be annihilated by the same bosonic oscillators and $S^a_r$
for $r>0$. As argued in
sec.(\ref{ss:R}) the fermionic zero modes do not contribute to the
hamiltonian in the R sector. The non-zero mode parts in the R and NS
sectors are given by,
\bea
2\alpha' p_- H_F^R = \sum_{n>0}w_n S^a_{-n} S^a_n ~, \quad
2\alpha' p_- H_F^{NS} = \sum_{r>0} w_r S^a_{-r} S^a_r + h_0(m)~,
\eea
where,
\bea
w_k = \sqrt{k^2 + \lt(m\over 2\pi \rt)^2}~,
\eea
with $k$ being either integer or half-odd-integer. The fermionic
oscillators satisfy the relevant anti-commutators of
eqs.(\ref{anti-comm-R}, \ref{anti-comm-NS}) with $\beta=i/2\pi$.
The zero point energy for the non-zero modes cancel between bosons and
fermions in the R sector. However, this does not happen in the NS
sector where it is given by,
\bea
h_0(m) &=& 4\lt( \sum_{n>0} w_n - \sum_{r>0} w_r  \rt) ~.
\label{h0-form}
\eea
This contains divergent sums and usually one regularises this
following standard methods. This has been done in appendix
\ref{a:regularization}
and the result is a non-trivial
function of $m$ involving
certain modified Bessel functions. But as mentioned before, unlike
in the case of flat space we do not yet have an independent physical
understanding of this result. It would be interesting to explore
this issue in future.

\sectiono{Generalisation}
\label{s:generalization}

Plane wave backgrounds are special cases of a more general backgrounds
called pp-waves which admit a covariantly constant null Killing
symmetry. In all such backgrounds the standard light-cone gauge can be
fixed \cite{lc}. Quite generically the world-sheet theories in
light-cone gauge involve a mass scale set by $p_-$ \cite{maldacena02}
such that the light-cone Hamiltonian takes the following form:
\bea
p_- H_{pp} = p_- H_{flat} + {\cal I}(p_-)~,
\eea
where ${\cal I}(p_-)$ is a collection of interaction terms
representing the deformation from flat space. This interaction term
depends on $p_-$ in such a way that,
\bea
\lim_{p_-\to 0} {\cal I} (p_-) \to 0~.
\label{I-limit}
\eea
This is clearly the case for the RR plane wave background, considered
in detail in this paper. This is also true for some of the cases
explicitly studied in the literature
\cite{generalization, papadopoulos02}.
This raises the following question:
\begin{center}
{\it Does the universality hold in all the pp-wave backgrounds?}
\end{center}
This question will be addressed in \cite{us}. The answer turns out to
be yes (which implies that the space-time action will be evaluated to
be same around the pp-wave backgrounds) only when all the scalars
(for example dilaton) that might be
switched on in the background take the same
profile\footnote{For the model studied in \cite{papadopoulos02} the
interaction is of special type such that eq.(\ref{I-limit}) does not
hold in light-cone gauge. But the universality might still work as the
interaction term does drop off in the linearised equation of motion in
space-time for configurations which do not depend on $x^-$ (see
eq.(5.19) in \cite{papadopoulos02}).}. Notice that
dilaton is constant in both the flat and RR plane wave backgrounds
studied here. This can further be generalised to non-trivial
transverse spaces or to cases where the transverse space is
consistently replaced by a conformal field theory.

\medskip
\centerline{\bf Acknowledgement}
\noindent
I am thankful to Sumit R. Das, Matthias R. Gaberdiel, Michael B. Green,
Martin Kruczenski, Juan Maldacena, Ignacio Navarro, Alfred D. Shapere and
Aninda Sinha for useful discussion. A preliminary version of this work
was presented at EFI, University of Chicago; YITP, Stony Brook,
Kentucky University, Great Lakes Strings Conference 2006, MCTP,
University of Michigan in March 2006 and Purdue University in October
2006. I thank the High Energy Physics Theory groups at the above
mentioned places and the organisers of the Michigan conference for
their interest. I am also grateful to Michael B. Green for reading an
earlier version of the manuscript and a following very useful
discussion and Sumit R. Das and Alfred D. Shapere for their continuous
encouragement and for supporting visits to Kentucky University where
parts of this work were done. This work was financially supported
by PPARC.

\appendix

\sectiono{Relevant features of even D-branes in RR plane wave}
\label{a:even}

Here we shall briefly review certain features of the open
strings on even D-branes in RR plane wave background that are relevant
to the R-sector zero mode problem for the non-BPS D-branes discussed in
this paper.
Because of the mass term the zero modes can either be $\tau$
or $\sig$ dependent. This leads
to two kinds of zero mode structures for the even D-barnes. The $\tau$
dependent zero modes end up giving a non-zero contribution to the
light-cone hamiltonian which breaks the degeneracy of the
$8_{\hbox{v}}$ and $8_{\hbox{c}}$ ground states. These D-branes have
been called Class I \cite{bergman02} or $D_-$ \cite{skenderis02}
branes in the literature. This lifting of the degeneracy can be
understood in the following way. All these
D-branes are either D$7$ or their dimensional reductions. The
orientation of such a D-brane is usually denoted as $(n,n\pm 2)$,
which implies that in addition to $x^{\pm}$ the brane is extended
along $n$ directions out of $x^1,\cdots ,x^4$ and $(n\pm 2)$
directions out of $x^5, \cdots , x^8$. Notice that the maximum
worldvolume dimensionality of such a D-brane is $8$ corresponding to a
D$7$-brane. A D$7$-brane is special in this background in the
following sense. Its eight dimensional worldvolume contains a
Chern-Simons (CS) coupling with the background RR $4$-form potential
which is quadratic in the worldvolume field strength
\cite{dabholkar02}. Hence this
introduces certain extra term in the linearised equations of motion
for the transverse components of the worldvolume vector potential
which is responsible for the lifting of degeneracy of
the open string ground states in $8_{\hbox{v}}$.
The supersymmetric version of this explains
why the degeneracy of the $8_{\hbox{c}}$ states should also be
broken. None of the other even D-branes (called Class II or $D_+$
branes) contains such a CS coupling. The open string zero modes for
such a D-brane is $\sig$ dependent which do not contribute to the
light-cone hamiltonian preserving the degeneracy of the $16$ ground
states.

\sectiono{Open String Boundary Conditions in Green-Schwarz Formalism}
\label{a:openbc}

Here we shall show that the open string boundary conditions for the
BPS and non-BPS D-branes of type IIB string theory in flat background
that can be analysed using kappa gauge fixed GS action are
also valid boundary conditions in the RR plane-wave background. Below
we shall first review the flat background where our discussion will be
both in the pre and post kappa gauge fixed theory. Later we shall show
why the kappa gauge fixed boundary conditions are also valid in RR
plane wave.

\subsectiono{Flat Background}
\label{as:flat}

After gauge fixing the world-sheet metric:
$g_{a b}=\eta_{a b}$ the GS world-sheet
action takes the following form:
\bea
S &\propto& S_1 + S_2~, \cr
S_1 &=&
-{1\over2} \int d^2\sigma \eta^{ab} \eta_{\mu \nu} \lt(\del_aX^{\mu} -
i(\theta^A \gb^{\mu} \del_a\theta^A)\rt) \lt(\del_b X^{\nu} -
i(\theta^B\gb^{\nu} \del_b\theta^B)\rt)~, \cr
S_2 &=&  \int d^2
\sigma \epsilon^{ab}\eta_{\mu \nu} \lt[-i\del_a
X^{\mu}\lt\{(\theta^1\gb^{\nu}\del_b \theta^1)-(\theta^2\gb^{\nu}\del_b
\theta^2) \rt\} +(\theta^1\gb^{\mu}\del_a \theta^1)
(\theta^2\gb^{\nu}\del_b \theta^2) \rt]~, \cr &&
\label{S-GS-flat}
\eea
where $a, b =0,1$, $\sigma^0=\tau$, $\sigma^1=\sigma$, $\epsilon^{ab}$ is the
two dimensional Levi-Civita symbol with $\epsilon^{01}=1$. $A=1,2$ and
we follow the convention of \cite{ps1} for the $10$-dimensional spinor
and the gamma matrices. The boundary terms obtained upon varying the
basic fields are given by,
\bea
\delta S_{boundary} &\propto& \int d\tau \lt[ - T_1 + i T_2 -
  T_3 + T_4+iT_5\rt]_0^{\pi} ~,
\label{deltaS-GS-flat}
\eea
where $[A]_0^{\pi} = A(\tau, \sigma=\pi) - A(\tau, \sigma=0)$ and
\bea
T_1 &=& \delta X^{\mu} \del_{\sigma} X_{\mu}~, \cr
T_2 &=& \del_{\sigma} X^{\mu} \lt\{ (\th^1\gb_{\mu}\delta \th^1) +
(\th^2\gb_{\mu} \delta \th^2) \rt\} - \del_{\tau} X^{\mu}
\lt\{(\th^1\gb_{\mu}\delta \th^1) -(\th^2\gb_{\mu}\delta \th^2)
  \rt\} ~, \cr
T_3 &=& (\th^1\gb_{\mu}\delta \th^1)(\th^2\gb^{\mu}\del_{\tau}\th^2) -
(\th^2\gb^{\mu}\delta \th^2)(\th^1\gb_{\mu}\del_{\tau}\th^1) ~, \cr
T_4 &= &  \lt\{(\th^1\gb_{\mu}\delta \th^1)+(\th^2\gb_{\mu}\delta \th^2)
  \rt\} \lt\{(\th^1\gb^{\mu}\del_{\sigma}\th^1) +
(\th^2\gb^{\mu}\del_{\sigma}\th^2) \rt\} ~, \cr
T_5 &=& \delta X^{\mu} \lt\{ (\th^1 \gb_{\mu}\del_{\sigma} \th^1) +
(\th^2\gb_{\mu}\del_{\sigma}\th^2)\rt\}
+ \delta X^{\mu} \lt\{ (\th^1 \gb_{\mu}\del_{\tau} \th^1) -
(\th^2\gb_{\mu}\del_{\tau}\th^2)\rt\}~.
\label{T12345}
\eea
Components of the stress-tensor are given by,
\bea
T_{\tau \tau} &=& T_{\sigma \sigma} = {1\over 2} \lt(\del_{\tau}X^{\mu}
\del_{\tau}X_{\mu} + \del_{\sigma}X^{\mu} \del_{\sigma}X_{\mu} \rt)
+i\del_{\tau} X_{\mu}\sum_A \lt(\th^A\gb^{\mu}\del_{\tau}\th^A \rt) +
i\del_{\sigma} \sum_A \lt(\th^A\gb^{\mu} \del_{\sigma}\th^A \rt)~, \cr
T_{\tau \sigma} &=& \del_{\tau}X^{\mu} \del_{\sigma}X_{\mu} + i
\del_{\tau}X_{\mu} \sum_A\lt(\th^A\gb^{\mu} \del_{\sigma}\th^A \rt)
+ i \del_{\sigma}X_{\mu} \sum_A \lt(\th^A \gb^{\mu} \del_{\tau}\th^A \rt)~.
\label{Tab}
\eea
Although the Virasoro constraint $T_{\tau \tau}=0$ imposes restriction
on the physical sates, in order for the energy eigenvalue to be
independent of $\tau$ we need to satisfy identically,
\bea
T_{\tau \sigma}=0~, \quad \hbox{at } \sigma = 0, \pi~.
\label{Ttausigma}
\eea
We shall now discuss open string boundary conditions corresponding to
BPS and non-BPS D-branes for which the variation in
(\ref{deltaS-GS-flat}) vanishes and eq.(\ref{Ttausigma}) is satisfied
identically. The standard Neumann and Dirichlet
boundary conditions for the space-time vectors are given by,
\bea
\hbox{Neumann: } &&
\del_{\sigma} X^{\mu}(\tau, \sigma) = 0~, \hbox{ at } \sigma = 0, \pi~, \cr
\hbox{Dirichlet: } && \delta X^{\mu}(\tau, \sigma) = 0 \hbox{ or }
\del_{\tau}X^{\mu}(\tau, \sigma) = 0~, \hbox{at } \sigma = 0, \pi~.
\label{NDbc-flat}
\eea
Due to this $T_1$ vanishes in equation (\ref{deltaS-GS-flat}). Boundary
conditions for the space-time spinors are different for BPS
and non-BPS D-branes which will be discussed separately below:

\begin{center}
{\bf \it BPS D-Branes}
\end{center}
For a BPS D-brane the relevant boundary condition is given by
\cite{green96, lambert99},
\bea
\th^1(\tau, \sigma) &=& M^S \th^2(\tau, \sigma)~, \quad \hbox{at }
\sigma =0,\pi ~,\cr
M^S &=& \g^{\mu_1 \cdots \mu_{p+1}}~,
\label{BPSbc-flat}
\eea
where $\mu_1, \cdots \mu_{p+1}$ correspond to the world-volume
directions of a D$p$-brane ($p$ odd). Boundary conditions involving
derivatives read,
\bea
\del_{\tau} \th^1(\tau, \sigma) = M^S \del_{\tau}
\th^2(\tau, \sigma)~, \quad
\del_{\sigma} \th^1(\tau, \sigma) = - M^S \del_{\sigma}
\th^2(\tau,\sigma)~, \quad \hbox{at } \sigma =0, \pi~.
\label{BPSbc-derivative}
\eea
One way to see the origin of the sign in the second equation is the
following. Define a spinor $\Theta(\tau, \sigma)$ on the cylinder:
$-\infty \leq \tau \leq \infty$, $0\leq \sigma \leq 2\pi$ by the
standard doubling trick:
\bea
\Theta(\tau, \sigma) = \lt\{ \begin{array}{ll}
\th^1(\tau,\sigma)~, & 0\leq \sigma \leq \pi ~, \cr
M^S \th^2(\tau, 2\pi -\sigma) ~, & \pi \leq \sigma \leq 2\pi ~.
\end{array}\rt.
\label{BPS-doubling}
\eea
Then for small $\epsilon$,
\bea
\del_{\sigma}\Theta(\tau, \sigma)|_{\sigma=\pi -\epsilon} =
\del_{\sigma}\theta^1(\tau, \sigma)|_{\sigma=\pi -\epsilon}~.
\label{eps1}
\eea
On the other hand,
\bea
\del_{\sigma}\Theta(\tau, \sigma)|_{\sigma=\pi +\epsilon} &=&
M^S \del_{\sigma}\theta^2(\tau, 2\pi -\sigma)|_{\sigma=\pi +\epsilon}~,
\cr
&=& -M^S \del_{\tilde \sigma}
\th^2(\tau, \tilde \sigma)|_{\tilde \sigma = \pi -\epsilon}~.
\label{eps2}
\eea
Taking the limit $\epsilon \to 0$ on the right hand sides of
eqs.(\ref{eps1}), (\ref{eps2}) and equating them one gets the second equation
in (\ref{BPSbc-derivative}). Using the boundary conditions
(\ref{BPSbc-flat}, \ref{BPSbc-derivative}) and
the relations,
\bea
(M^S)^T \gb^{\mu} M^S &=& - (M^V)^{\mu}_{~\nu} \gb^{\nu}~,
\label{MSTgbMS}
\eea
\bea
\lt.
\begin{array}{l}
\lt[\delta^{\mu}_{~\nu} - (M^V)^{\mu}_{~\nu}\rt] \del_{\sigma}X^{\nu}
=0~, \cr
\lt[\delta^{\mu}_{~\nu} + (M^V)^{\mu}_{~\nu}\rt] \del_{\tau}X^{\nu}
=0~, \hbox{ or }
\lt[\delta^{\mu}_{~\nu} + (M^V)^{\mu}_{~\nu}\rt] \delta X^{\nu}
=0~,
\end{array} \rt\} \hbox{at } \sigma =0, \pi ~,
\label{relation}
\eea
where,
\bea
(M^V)^{\mu}_{~\nu} &=& \lt\{ \begin{array}{rl}
-\delta^{\mu}_{~\nu}~, & \hbox{for } X^{\mu} \hbox{ Neumann}~, \cr
\delta^{\mu}_{~\nu}~, & \hbox{for } X^{\mu} \hbox{ Dirichlet}~.
\end{array} \rt.
\label{MV}
\eea
one can show that all the rest of the terms in the variation
(\ref{deltaS-GS-flat}) vanish. Similarly using the boundary conditions
(\ref{NDbc-flat}), (\ref{BPSbc-flat}) and (\ref{BPSbc-derivative}) and
the relation (\ref{relation}) one can show that eq.(\ref{Ttausigma})
is satisfied identically.

\begin{center}
{\bf \it Non-BPS D-Branes}
\end{center}
For non-BPS D-branes we propose, following \cite{ps1}, the following
bi-local boundary condition for the space-time spinors:
\bea
\th^{1\alpha}(\tau, \sigma) \th^{1\beta}(\tau',\sigma) = {\cal
  M}^{\alpha \beta}_{\g \delta}
\th^{2\g}(\tau, \sigma) \th^{2\delta}(\tau',\sigma)~, \quad \hbox{at }
\sigma = 0, \pi~,
\label{nBPSbc-flat}
\eea
where,
\bea
{\cal M}^{\alpha \beta}_{\g \delta} &=& -\lt[{1\over 16}
\g^{\alpha \beta}_{\mu} (M^V)^{\mu}_{~\nu} \gb^{\nu}_{\g \delta} +
{1\over 16 \times 3!} \g^{\alpha \beta}_{\mu_1 \cdots \mu_3}
(M^V)^{\mu_1}_{~\nu_1} \cdots
(M^V)^{\mu_3}_{~\nu_3} \gb^{\nu_1\cdots \nu_3}_{\g \delta} \rt. \cr
&& \lt. + {1\over 16\times 5!} \sum_{\mu_1, \cdots, \mu_5\in {\cal K}^{(5)}}
\g^{\alpha \beta}_{\mu_1 \cdots \mu_5} (M^V)^{\mu_1}_{~\nu_1}\cdots
(M^V)^{\mu_5}_{~\nu_5} \gb^{\nu_1\cdots \nu_5}_{\g \delta}\rt]~.
\label{calM}
\eea
The definition of the set ${\cal K}^{(5)}$ can be found in
\cite{ps2} (see discussion below eq.(2.3) in \cite{ps2}). We shall now
argue that all the terms $[T_2]_0^{\pi}, \cdots, [T_5]_0^{\pi}$
on the right hand side of eq.(\ref{deltaS-GS-flat})
vanish. Considering the term $T_2$ one finds,
\bea
[T_2]_0^{\pi} = \lt[ \lt\{ \del_{\sigma} X_{\mu} \lt(\delta^{\mu}_{~\nu} -
(M^V)^{\mu}_{~\nu} \rt) - \del_{\tau} X_{\mu} \lt(\delta^{\mu}_{~\nu} +
(M^V)^{\mu}_{~\nu} \rt) \rt\} (\th^2 \gb^{\nu} \delta \th^2)\rt]_0^{\pi} = 0~,
\eea
where we have used
\bea
\lt(\th^1(\tau, \sigma)\gb^{\mu}\th^1(\tau',\sigma) \rt) +
M^{\mu}_{~\nu} \lt(\th^2(\tau, \sigma) \gb^{\nu}\th^2(\tau',\sigma)
\rt) =0~, \quad \hbox{at } \sigma =0,\pi~,
\label{thmu-flat}
\eea
which can be obtained from the boundary condition
(\ref{nBPSbc-flat}). To treat the rest of the terms we need to know
the boundary conditions which involve variation and derivatives of the
space-time spinors. Since the boundary condition (\ref{thmu-flat}) is
bi-local we can vary the fields at the two points independently. This
gives,
\bea
\lt(\th^1(\tau, \sigma)\gb^{\mu}\delta \th^1(\tau',\sigma) \rt) +
M^{\mu}_{~\nu} \lt(\th^2(\tau, \sigma) \gb^{\nu} \delta \th^2(\tau',\sigma)
\rt) =0~, \quad \hbox{at } \sigma =0,\pi~.
\label{thmu-flat-vary}
\eea
One can also apply $\tau$-derivatives at the two points
independently. Therefore,
\bea
\lt(\th^1(\tau, \sigma)\gb^{\mu}\del_{\tau'} \th^1(\tau',\sigma) \rt) +
M^{\mu}_{~\nu} \lt(\th^2(\tau, \sigma) \gb^{\nu}
\del_{\tau'} \th^2(\tau',\sigma) \rt) =0~, \quad \hbox{at } \sigma =0,\pi~.
\label{thmu-flat-tderivative}
\eea
Using these it is straightforward to show that $[T_3]_0^{\pi}=0$. To
obtain the boundary condition involving $\sigma$-derivative we first
adopt a suitable doubling trick following \cite{nbps2}:
\bea
\Theta^{1\alpha}(\tau, \sigma) \Theta^{1\beta}(\tau', \sigma') = \lt \{
\begin{array}{ll}
\th^{1\alpha}(\tau,\sigma) \th^{1\beta}(\tau',\sigma') & 0\leq \sigma,
\sigma' \leq \pi~, \cr
{\cal M}^{\alpha \beta}_{\g \delta} \th^{2\g}(\tau,2\pi-\sigma)
\th^{2\delta}(\tau',2\pi-\sigma') & \pi \leq \sigma, \sigma' \leq 2\pi~.
\end{array} \rt.
\label{doublingTh}
\eea
Then using the similar reasoning as done in the BPS-case one can show:
\bea
\lt(\th^1(\tau, \sigma)\gb^{\mu}\del_{\sigma} \th^1(\tau',\sigma) \rt) -
M^{\mu}_{~\nu} \lt(\th^2(\tau, \sigma) \gb^{\nu}
\del_{\sigma} \th^2(\tau',\sigma) \rt) =0~, \quad \hbox{at } \sigma =0,\pi~.
\label{thmu-flat-sderivative}
\eea
Using this, the boundary conditions (\ref{thmu-flat-tderivative}) and
(\ref{relation}) one can show $[T_4]_0^{\pi} = 0$ and
$[T_5]_0^{\pi}=0$. This shows that for the boundary conditions
(\ref{NDbc-flat}) and (\ref{nBPSbc-flat}) the boundary terms
(\ref{deltaS-GS-flat}) in the variation of the wold-sheet action are
zero. Also using eqs.(\ref{NDbc-flat}), (\ref{thmu-flat-tderivative}),
(\ref{thmu-flat-sderivative}) and (\ref{relation}) one can establish that
eq.(\ref{Ttausigma}) is satisfied identically.

\begin{center}
{\bf \it D-Branes in Kappa Gauge Fixed Theory}
\end{center}
If we fix the fermionic kappa symmetry by requiring:
\bea
\gb^+ \th^A = 0~, \Rightarrow \th^A = c \pmatrix{S^a \cr 0}~,
\label{kappa-fix}
\eea
then we have,
\bea
T_2 &=& \sqrt{2}c^2 \lt[\del_{\sigma}X^+
\lt\{ (S^1\delta S^1)+(S^2\delta S^2)\rt\}-\del_{\tau} X^+
\lt\{(S^1\delta S^1) - (S^2\delta S^2) \rt\}\rt]~, \cr
T_3&=& 0~, \cr
T_4&=& 0~, \cr
T_5&=& \sqrt{2}c^2\lt[\delta X^+
\lt\{(S^1\del_{\sigma}S^1)+(S^2\del_{\sigma}S^2) \rt\} + \delta X^+
\lt\{(S^1\del_{\tau}S^1)-(S^2\del_{\tau}S^2) \rt\}\rt]~,
\label{T2345-kappa}
\eea
and
\bea
T_{\tau \sigma} &=& \del_{\tau}X^+\del_{\sigma}X^-
+\del_{\sigma}X^+\del_{\tau}X^- +\del_{\tau}X^I\del_{\sigma}X^I \cr &&
 +\sqrt{2}ic^2\del_{\tau}X^+\sum_A(S^A\del_{\sigma}S^A) + \sqrt{2}ic^2
 \del_{\sigma}X^+ \sum_A(S^A\del_{\tau}S^A)~.
\label{Ttausigma-kappa}
\eea
For all the D-branes that can be described in this gauge, both the
light-cone coordinates $X^{\pm}$ should satisfy Neumann boundary
condition\footnote{In order for the first term in the expression for
  $T_2$ in eqs.(\ref{T2345-kappa}) to vanish at the boundary we need
  $X^+$ to satisfy Neumann boundary condition. Then in order for the
  first term in eq.(\ref{Ttausigma-kappa}) to vanish at the boundary
  we need $X^-$ also to satisfy Neumann boundary condition.}:
\bea
\del_{\sigma} X^{\pm}(\tau, \sigma) = 0 ~, \quad \hbox{at } \sigma =0,\pi~.
\eea
Boundary condition for the space-time spinors for a BPS D-brane reads:
\bea
S^{1a}(\tau, \sigma) &=& {\cal M}^S_{ab} S^{2b}(\tau, \sigma) ~, \quad
\hbox{at } \sigma = 0, \pi~, \cr
{\cal M}^S_{ab} &=& \lt(\sigma^{I_1I_2\cdots I_{p-1}}\rt)_{ab}~,
\eea
where $X^{I_1}, \cdots , X^{I_{p-1}}$ are the Neumann directions
besides $X^{\pm}$. For a non-BPS D-brane one gets,
\bea
S^{1a}(\tau,\sigma) S^{1b}(\tau', \sigma) &=& {\cal M}^{ab}_{cd}
S^{2c}(\tau,\sigma) S^{2d}(\tau', \sigma)~,
\label{nBPSbc-lc}
\eea
where,
\bea
{\cal M}^{ab}_{cd} &=& {1\over 8} \delta_{ab}\delta_{cd} + {1\over 16}
\sum_{IJ} \lambda_I \lambda_J \sigma^{IJ}_{ab} \sigma^{IJ}_{cd}
+{1\over 192}\sum_{IJKL\in {\cal K}^{(4)}} \lambda_I
\lambda_J\lambda_K\lambda_L \sigma^{IJKL}_{ab} \sigma^{IJKL}_{cd}~.
\label{calM-lc}
\eea
As explained in \cite{nbps2}, the summation over the vector indices in
the last term is restricted and a definition for the set ${\cal
  K}^{(4)}$ can be found, for example, in \cite{ps2} (see discussion
below eq.(4.9)). $\lambda_I$ is the eigenvalue of $M^V$ in
eq.(\ref{MV}) along the light-cone transverse direction $x^I$.

\subsectiono{RR Plane Wave Background}
\label{as:RRplanewave}

In case of the RR plane wave background the boundary terms in
the variation of the world-sheet action read:
\bea
\delta S^{boundary}_{pw} = \delta S^{boundary} + \delta (\Delta S)~,
\label{deltaS-GS-pp}
\eea
where $\delta S^{boundary}$ is given in eq.(\ref{deltaS-GS-flat}) with
$T_1$ given in (\ref{T12345}) and the $T_2,\cdots, T_5$ given in
(\ref{T2345-kappa}) and,
\bea
\delta (\Delta S) \propto \int d\tau \lt[\mu^2 \vec X^2 \delta X^+
  \del_{\sigma}X^+ + 4\sqrt{2}ic^2\mu \delta
  X^+\del_{\sigma}X^+(S^1\Pi S^2) \rt]_0^{\pi}~.
\eea
The off-diagonal component of the stress-tensor is given by,
\bea
(T_{pw})_{\tau \sigma} = T_{\tau \sigma} + \Delta T_{\tau \sigma}~,
\eea
where $T_{\tau \sigma}$ is given in eq.(\ref{Ttausigma-kappa}) and,
\bea
\Delta T_{\tau \sigma} = -\mu^2 \vec X^2 \del_{\tau}X^+
\del_{\sigma}X^+ - 4\sqrt{2}ic^2\mu \del_{\tau}X^+ \del_{\sigma}X^+
(S^1\Pi S^2)~.
\eea
Notice that all the terms in $\delta (\Delta S)$ and $\Delta T_{\tau
  \sigma}$ contain $\del_{\sigma} X^+$. Since $X^+$ satisfies Neumann
boundary condition both $\delta (\Delta S)$ and $\Delta T_{\tau
  \sigma}$ at the boundary are zero. This shows
that all the open string boundary conditions in flat space are also
valid boundary conditions in the RR plane wave.

\sectiono{Identities}
\label{a:identities}

Here we present some useful identities that are required for various
computations done in this paper. Using the following completeness relation,
\bea
{1\over 8} \delta_{ab} \delta_{ef} + {1\over 16} \sum_{I,J}
\sigma^{IJ}_{ab}\sigma^{IJ}_{ef} +{1\over 384} \sum_{I,J,K,L}
\sigma^{IJKL}_{ab} \sigma^{IJKL}_{ef} = \delta_{ae}\delta_{bf} ~,
\label{completeness}
\eea
one can show,
\bea
\M^{ab}_{cd} \M^{cd}_{ef} = \delta_{ae} \delta_{bf}~.
\label{inverse}
\eea
We also have,
\bea
\M^{ab}_{cd} = \M^{ba}_{dc}~,
\label{transpose}
\eea
\bea
\Pi^T = \Pi~, \quad \Pi \sigma^{IJ\cdots} = (\beta_I \beta_J \cdots)
\sigma^{IJ\cdots} \Pi~,
\label{Pi-prop}
\eea
where in the last equation of (\ref{Pi-prop}) product of an even number
of sigma matrices has been considered and,
\bea
\beta_I = \lt\{ \begin{array}{ll}
1 & \hbox{if } I \in \{1,2,3,4\} \\
-1 & \hbox{otherwise}~.
\end{array} \rt.
\label{beta}
\eea
Using the above relations one can prove the following identities:
\bea
\lt[ \lt(\Pi^{\epsilon_1} \over \Pi^{\epsilon_2}\rt) \M
  \lt(\Pi^{\epsilon_3}\over \Pi^{\epsilon_4}\rt) \rt]^{ab}_{cd} =
\lt[ \lt(\Pi^{1-\epsilon_1} \over \Pi^{1-\epsilon_2}\rt) \M
  \lt(\Pi^{1-\epsilon_3}\over \Pi^{1-\epsilon_4}\rt) \rt]^{ab}_{cd}~,
\eea
\bea
\lt[ \lt(\Pi^{\epsilon_1} \over \Pi^{\epsilon_2}\rt) \M
  \lt(\Pi^{\epsilon_3}\over \Pi^{\epsilon_4}\rt) \rt]^{ab}_{cd} =
\lt[ \lt(\Pi^{\epsilon_3} \over \Pi^{\epsilon_4}\rt) \M
  \lt(\Pi^{\epsilon_1}\over \Pi^{\epsilon_2}\rt) \rt]^{ba}_{dc} ~,
\eea
\bea
\lt[ \lt(\Pi^{\epsilon_1} \over \Pi^{\epsilon_2}\rt) \M
  \lt(\Pi^{\epsilon_3}\over \Pi^{\epsilon_4}\rt) \rt]^{ab}_{cd}
\lt[ \lt(\Pi^{\epsilon_2} \over \Pi^{\epsilon_1}\rt) \M
  \lt(\Pi^{\epsilon_4}\over \Pi^{\epsilon_3}\rt) \rt]^{cd}_{ef}
=\delta_{ae} \delta_{bf} ~,
\eea
where $\epsilon_i=0,1$ which correspond to not having and having a
factor of $\Pi$ respectively.

\sectiono{On the resolution of the zero-mode problem}
\label{a:zero}
In section \ref{ss:R} the R-sector zero-mode spectrum in
eqs.(\ref{zero-mode-spectrum}) has been obtained by requiring consistency
with the relevant worldvolume interpretation. But it should also be
possible to get the same result from the open string theory. As mentioned
earlier it is not clear how to arrive at the result in
eqs.(\ref{zero-mode-spectrum}) starting from the algebraic structure
as given in eqs.(\ref{StSt}, \ref{anti-comm-SSt}) and (\ref{D-antisymm}).
Here we shall discuss certain curious features of the problem that might
be relevant for the actual resolution. Notice that the spectrum in
eqs.(\ref{zero-mode-spectrum}) is exactly same as that in flat space
where the definition of the ground states is known. If we follow the same
definition then we have\footnote{Without any loss of generality we have chosen
$\beta=i/2\pi$ to avoid clutter in various expressions.}:
\bea
S^a_0 |I\ra = {1\over \sqrt{2}} \sig^I_{a\dot a} |\dot a\ra~, \quad
S^a_0 |\dot a\ra = {1\over \sqrt{2}} \sig^I_{a\dot a} |I\ra~.
\label{Iadot}
\eea
Then the problem will be to compute:
\bea
z^I_{a\dot a} := \la I| \St^a_0 |\dot a\ra ~.
\label{z}
\eea
This can be done very easily: using eq.(\ref{H0}) and $\la \dot a
|H_0|\dot b \ra =0$ one gets,
\bea
\sig^I_{a\dot a} z^I_{a\dot b} = 0~.
\label{sig-z}
\eea
Then expanding $z^I_{a\dot a}$ in complete basis of odd rank $SO(8)$
gamma matrices:
\bea
z^I_{a\dot a} = c\sig^I_{a\dot a} + c_{JK} \sig^{IJK}_{a\dot a}~,
\label{expand-z}
\eea
and substituting into eq.(\ref{sig-z}) one gets: $c=0$, $c_{IJ}=0$ implying,
\bea
z^I_{a\dot a} = 0~.
\label{z-0}
\eea

There is also a way to get the same result using the existing
algebraic structure
without having any prior knowledge of $H_0$ eigenvalues. Using the
first set of anti-commutators in eqs.(\ref{anti-comm-SSt}) one
obtains,
\bea
\lt( \St_0 \St_0 \rt) = \lt( S_0 S_0 \rt) = 4 ~, \quad
\lt(\St \sig^{IJKL} \St \rt) = \lt(S \sig^{IJKL} S \rt) = 0~, \quad
\forall~ I,J,K,L~.
\eea
Using these equations and the first relation in (\ref{StSt}) one can show,
\bea
\St^a_0 = {1\over 56} \sum_{IJ} \hat \lam_{IJ} \lt(S\sig^{IJ}S\rt)
\lt(\sig^{IJ}\St\rt)^a~.
\label{linear-St}
\eea
In this particular discussion we shall write the summations explicitly
instead of following the repeated-index convention for them. At this stage we
introduce the ground states as in eqs.(\ref{Iadot}). Then using
eq.(\ref{linear-St}) one obtains:
\bea
z^K_{a\dot a} = {1\over 56} \sum_{IJLb} \hat \lam_{IJ} \la
K|\lt(S\sig^{IJ}S\rt)|L\ra \sig^{IJ}_{ab} z^L_{b\dot a}~.
\label{eq-z1}
\eea
Using the result: $\la K|\lt(S\sig^{IJ}S\rt)|L\ra =
4(\delta_{IK}\delta_{JL} -\delta_{IL}\delta_{JK})$ one finds,
\bea
\sum_{Jb} \lt(7 \delta^{IJ} \delta_{ab} -\hat \lam_{IJ} \sig^{IJ}_{ab}
\rt) z^J_{b\dot a} =0~.
\label{eq-z2}
\eea
Then expanding $z^I_{a\dot a}$ as in eq.(\ref{expand-z}) and substituting
this back into eq.(\ref{eq-z2}) one gets an equation whose right hand
side is zero and left hand side is expanded in terms of first and
third rank $SO(8)$ gamma matrices. We then set the coefficient of each
independent term to zero. For example, the coefficient of $\sig^I$ and
$\sig^J~ (J\neq I)$ are given
by $c\{8-(8-2d)\hat \lam_I \}$ and $-2c_{IJ}\{(8-2d)\hat \lam_I
-1-\hat \lam_{IJ}\}$ respectively, where $d+2=p+1$ is the dimension of
worldvolume.  This gives the same result as in eq.(\ref{z}).

Given this result one may use the completeness relation: $\la I|I\ra +
  \la \dot a|\dot a\ra=1$ to conclude that $\St^a_0$ does not appear
  as a non-trivial operator in the theory. Notice that for the even
  D-branes and also for the non-BPS D-branes in flat
  space\footnote{Open string spectrum for a non-BPS D-brane was
  obtained in \cite{nbps2} from the bi-local boundary condition
  without encountering the present problem. This is simply because in
flat background the fermions are massless and there is no mixing
between $S$ and $\St$ variables. Therefore all the $\St$ variables,
including the zero modes, are classically eliminated against the $S$
variables. Because
of this the operator $\St^a_0$ does not appear at all in the open
string theory.} $\St^a_0$ is completely eliminated classically so that
  it does not appear at all in the quantum theory. The above result
  seems to suggest that in our case $\St^a_0$ is eliminated in the
  full quantum theory.

We have argued before that $\D^{ab}$, as defined in
(\ref{anti-comm-SSt}), can not be a non-zero c-number, neither can it
be fixed in the present theory. Notice that the above result
explains this situation as it implies that an arbitrary matrix element
of $\D^{ab}$, which is potentially an operator, is always
zero. However it is not clear to us how to interpret, for example, the
non-zero anti-commutator of $\St^a_0$ from the present point of view.

\begin{center}
{\large \it An argument using bosonisation}
\end{center}

Here we shall attempt to understand the same result i.e. why
$\St^a_0$ may not appear as a non-trivial operator in the theory
once the ground states are defined as in eqs.(\ref{Iadot}) in a
different way using bosonisation. This approach is also not completely
understood to us, but the following arguments show that it may be worth
understanding this better.

Notice that the zero mode structure that we are
dealing with, namely the first equation in (\ref{StSt}), also
appears in the discussion of non-BPS D-branes in flat background.
Therefore it should be possible to understand these rules in that
context itself. We therefore consider the massless fermions for
simplicity for the present argument. From the mode expansion in the
bulk one gets,
\bea
S^a_0 = \oint {dz\over 2\pi i} z^{-1/2} S^a(z)~, \quad \St^a_0 = -
\oint {d\bar z \over 2\pi i} \bar z^{-1/2} \St^a(\bar z)~,
\label{SSt0-int}
\eea
where the coordinates $(z,\bar z)$ on the upper half plane (UHP) are
related to the coordinates $(\tau, \sig)$ on the strip (euclidean) as:
$z=e^{\tau + i\sig}$. For the operators in the bulk the contours
in eqs.(\ref{SSt0-int}) do not pass through the real line. To define
the above zero modes on the boundary one can go to the doubled surface
by using the relevant boundary condition. This boundary condition
needs to be linear as the right hand sides of eqs.(\ref{SSt0-int}) are
linear in GS fermions. We already know that the boundary condition is
actually quadratic. To see what kind of linear boundary conditions one
might expect one can first bosonise $S^a(\tau, \sigma)$ and
$\St^a(\tau, \sig)$ in terms of four left and right moving bosons:
$\phi^i(\tau , \sig)$ and $\tilde \phi^i(\tau ,\sig)$ respectively
with $i=1,\cdots , 4$ \cite{friedan85, kostelecky86}. Then using the relevant
boundary conditions relating $\phi$ and $\tilde \phi$
\footnote{Non-BPS D-branes were previously studied in \cite{nbps1} using such
  non-abelian bosonisation from the closed string point of
  view. Similar analysis can possibly be done in the open string
  picture to find boundary conditions for the bosons.}
and properly refermionising them back one will arrive at the desired
expression. This is a detailed task by itself and we shall not attempt
to do it explicitly here. But it seems plausible that the manifest
  $SO(8)$ covariance will require the final expression to be of the
  following form:
\bea
\St^a(\tau,\sig) = \widehat \M_{a\dot a} S^{\dot a}(\tau, \sig)~,
\quad S^a(\tau,\sig) = \widehat \M_{a\dot a} \St^{\dot a}(\tau,
\sig)~, \quad
\hbox{at } \sig =0, \pi~,
\label{linearbc}
\eea
where $\widehat \M = \sig^{I_1\cdots I_{(p-1)}}$ is the spinor
representation corresponding to the set of reflections given by
$\hat \lam_I$ in (\ref{lamhat}). Following \cite{gsw, witten84} one
would then conclude
that $S^{\dot a}(\tau, \sig)$ and $\St^{\dot a}(\tau,
\sig)$ are the spin fields that can be obtained by suitably bosonising
and refermionising $S^a(\tau,\sig)$ and $\St^a(\tau,\sig)$
respectively.

We shall now discuss the doubling trick in the present context. By
using this trick one usually reduces an arbitrary correlator on the UHP to a
holomorphic correlator on the full plane. This trick was suitably
developed in \cite{nbps2} to handle the bi-local boundary conditions
relevant to our case. It was shown that a correlator with an arbitrary
number of bulk insertions on the UHP, but without any boundary
insertions, can be mapped to a holomorphic correlator involving the
holomorphic triad: $\{{\cal S}^a(z), {\cal S}^{\dot a}(z), \Psi^I(z)
\}$. If the doubling trick works then the same should be true even in
presence of the boundary insertions. A suitable doubling that properly
implements this is the following,
\bea
{\cal S}^a(\tau,\sig) = \lt\{
\begin{array}{ll}
S^a(\tau,\sig) & 0\leq \sig \leq \pi ~,\cr
\widehat \M_{a\dot a}\St^{\dot a}(\tau, 2\pi -\sig) & \pi \leq \sig
\leq 2\pi ~.
\end{array} \rt.
\eea
\bea
{\cal S}^{\dot a}(\tau,\sig) = \lt\{
\begin{array}{ll}
S^{\dot a}(\tau,\sig) & 0\leq \sig \leq \pi ~,\cr
\widehat \M^{-1}_{\dot a a}\St^a(\tau, 2\pi -\sig) & \pi \leq \sig \leq 2\pi ~.
\end{array} \rt.
\eea
We can now express the zero modes on the boundary in the following way:
\bea
S^a_0 = \oint {dz\over 2\pi i} z^{-1/2} {\cal S}^a(z)~, \quad \St^a_0
= \widehat \M_{a\dot a} \oint {dz \over 2\pi i} z^{-1/2} {\cal
  S}^{\dot a}(z)~,
\label{SSt0-int-double}
\eea
where the contours are understood to be around points on the real line.

Given eq.(\ref{SSt0-int-double}) we can now proceed to compute the
matrix element in (\ref{z}):
\bea
z^I_{a\dot a} = \lim_{w\to0} {1\over w} \widehat \M_{a\dot
  b} \oint {dz\over 2\pi i} z^{-1/2}
\la \Psi^I(-1/w) {\cal S}^{\dot b}(z) {\cal S}^{\dot a}(0) \ra =0~,
\eea
as the holomorphic correlator itself is zero because of the momentum
conservation of the bosons. Another way to understand why
$\St^a_0$ does not appear as a non-trivial operator is the following:
Notice that the operators in (\ref{SSt0-int-double}) involve branch
cuts. For any sensible
computation these branch cuts need to be cancelled by other branch cuts
originating from operators in the neighbourhood. It turns out that
$\St^a_0$ is not generically well defined in this sense in the
neighbourhood of the ground states as defined in eqs.(\ref{Iadot}). For
example, let us consider the following matrix element: $\la \dot
b|\St^b_0 \St^a_0 |\dot a\ra$. Although the momentum conservation can
be satisfied in this case, branch cuts are not removed:
\bea
\la \dot b|\St^b_0 \St^a_0 |\dot a\ra
= \lim_{u\to 0}{1\over u} \widehat \M_{b\dot c} \widehat
\M_{a\dot d} \oint {dz\over 2\pi i} z^{-1/2} \oint
{dw \over 2\pi i} w^{-1/2} \la {\cal S}^{\dot b}(-1/u) {\cal S}^{\dot
  c}(z) {\cal S}^{\dot d}(w) {\cal S}^{\dot a}(0) \ra~.
\eea
Contracting the $w$-contour to zero and using the OPE:
${\cal S}^{\dot d}(w){\cal S}^{\dot a}(0) \sim
\delta_{\dot d \dot a}/w$ one finds that the branch cut in $w$ is not removed.

Although the above explanation may seem to be convincing, the
present approach has to be understood better because of a
potential caveat that we shall discuss now. If we have
a linear boundary condition like in (\ref{linearbc}) then it is
expected that the following bi-local boundary condition \cite{nbps2}
will be reproduced:
\bea
\St^a(\tau, \sig) \St^b(\tau', \sig) = \widehat \M^{ab}_{cd}
S^c(\tau,\sig) S^d(\tau',\sig)~, \quad \hbox{at } \sig=0,\pi ~,
\eea
where $\widehat {\cal M}^{ab}_{cd}$ is defined in eq.(\ref{calMhat}).
This will be satisfied if the following equation holds as an equality of OPE's:
\bea
\widehat \M_{a\dot a} \widehat \M_{b\dot b} {\cal S}^{\dot a}(z) {\cal
  S}^{\dot b}(w) &=&\widehat \M^{ab}_{cd} {\cal S}^c(z) {\cal
  S}^d(w)~.
\label{show-OPE}
\eea
This is almost satisfied, but not quite because of the subtlety
involving fourth rank tensors. The fourth rank tensor constructed out
of ${\cal S}^a(z)$ is always self-dual whereas that constructed out
of ${\cal S}^{\dot a}(z)$ is always anti-self-dual. To see this more
explicitly notice that the relevant OPEs can be computed using
bosonisation as in, for example, \cite{kostelecky86}. The final results can
be expanded in terms of the complete basis constructed out of $SO(8)$
gamma matrices. For the conjugate spinors one gets \cite{kostelecky86},
\bea
{\cal S}^{\dot a}(z) {\cal S}^{\dot b}(w) &=& \delta_{\dot a\dot b}
\sum_{n\geq -1} c^{(0)}_n (z-w)^n {\cal A}^{(0)}_n(\Psi(w)) + \bar
\sig^{IJ}_{\dot a\dot b} \sum_{n\geq 0}c^{(2)}_n(z-w)^n {\cal
  A}^{(2)IJ}_n (\Psi(w)) \cr
&& + \bar \sig^{IJKL}_{\dot a\dot b} \sum_{n\geq 1}c^{(4)}_n(z-w)^n
    {\cal A}^{(4)IJKL}_n(\Psi(w))~, \cr
&=& \delta_{\dot a\dot b}
\sum_{n\geq -1} c^{(0)}_n (z-w)^n {\cal A}^{(0)}_n(\Psi(w)) + \bar
\sig^{IJ}_{\dot a\dot b} \sum_{n\geq 0}c^{(2)}_n(z-w)^n {\cal
  A}^{(2)IJ}_n (\Psi(w)) \cr
&& + \sum_{I,J,K,L \in \K^{(4)}}\bar \sig^{IJKL}_{\dot a\dot b}
\sum_{n\geq 1}c^{(4)}_n(z-w)^n {\cal A}^{(4,-)IJKL}_n(\Psi(w))~,
\label{cc-OPE}
\eea
where ${\cal A}^{(0)}_n$, ${\cal A}^{(2)}_n$ and ${\cal A}^{(4)}_n$
are anti-symmetric tensors of rank as indicated by the superscripts
constructed using $\Psi^I(w)$ and its derivatives. Notice that in the
fist step the summation over the vector indices on the fourth rank
tensors is free. But all $\bar \sig^{IJKL}$ matrices are not
independent. In the second step we have restricted this summation over
only the independent ones\footnote{This requires us to choose which ones we
  want to keep as independent. This is reflected in the choice of
the set $\K^{(4)}$ that also appears in the bi-local boundary
condition in (\ref{nBPSbc-lc}, \ref{calM-lc}).} using the
anti-self-duality of the $\bar \sig^{IJKL}$ matrices.
${\cal A}^{(4,-)IJKL}_n= {\cal A}^{(4)IJKL}_n -
{1\over 4!} \epsilon^{IJKLMNPQ} {\cal A}^{(4)MNPQ}_n$ is
therefore the anti-self-dual component of ${\cal A}^{(4)IJKL}_n(\Psi(w))$.
Then using the following relations:
\bea
\lt(\widehat \M \widehat \M^T\rt)_{ab} = \delta_{ab}~, \quad
\lt(\widehat \M\bar \sig^{IJ}\widehat \M^T\rt)_{ab}=\hat \lam_{IJ}
\sig^{IJ}_{ab}~,\quad \lt(\widehat \M\bar \sig^{IJKL}\widehat
\M^T\rt)_{ab} = \hat \lam_{IJKL} \sig^{IJKL}_{ab}~,
\label{sigbar-sig}
\eea
one finds for the left hand side of eq(\ref{show-OPE}),
\bea
\widehat \M_{a\dot a} \widehat \M_{b\dot b} {\cal S}^{\dot a}(z) {\cal
  S}^{\dot b}(w) &=&
\delta_{ab}
\sum_{n\geq -1} c^{(0)}_n (z-w)^n {\cal A}^{(0)}_n(\Psi(w)) \cr
&& + \hat \lam_{IJ} \sig^{IJ}_{ab} \sum_{n\geq 0}c^{(2)}_n(z-w)^n {\cal
  A}^{(2)IJ}_n (\Psi(w)) \cr
&& + \sum_{I,J,K,L \in \K_{(4)}}\hat \lam_{IJKL} \sig^{IJKL}_{ab}
\sum_{n\geq 1}c^{(4)}_n(z-w)^n {\cal A}^{(4,-)IJKL}_n(\Psi(w))~,  \cr &&
\label{lhs-show-OPE}
\eea
The OPE for the spinors is very similar to that in eq.(\ref{cc-OPE})
\cite{kostelecky86}:
\bea
{\cal S}^a(z) {\cal S}^b(w) &=& \delta_{ab} \sum_{n\geq -1} c^{(0)}_n
(z-w)^n {\cal A}^{(0)}_n(\Psi(w)) + \sig^{IJ}_{ab} \sum_{n\geq
  0}c^{(2)}_n(z-w)^n {\cal A}^{(2)IJ}_n (\Psi(w)) \cr
&& + \sum_{I,J,K,L \in \K_{(4)}}
\sig^{IJKL}_{ab} \sum_{n\geq 1}c^{(4)}_n(z-w)^n {\cal
  A}^{(4,+)IJKL}_n(\Psi(w))~,
\label{ss-OPE}
\eea
where in the last term the self-dual component:
${\cal A}^{(4,+)IJKL}_n= {\cal A}^{(4)IJKL}_n +
{1\over 4!} \epsilon^{IJKLMNPQ} {\cal A}^{(4)MNPQ}_n$ appears. Then
using various traces of the gamma matrices one shows:
\bea
\sum_{n\geq -1} c^{(0)}_n
(z-w)^n {\cal A}^{(0)}_n(\Psi(w)) &=& {1\over 8} {\cal S}^a(z) {\cal
  S}^a(w)~, \cr
\sum_{n\geq
  0}c^{(2)}_n(z-w)^n {\cal A}^{(2)IJ}_n (\Psi(w)) &=& {1\over 16}
\sig^{IJ}_{ab} {\cal S}^a(z) {\cal S}^b(w)~, \cr
\sum_{n\geq 1}c^{(4)}_n(z-w)^n {\cal
  A}^{(4,+)IJKL}_n(\Psi(w)) &=& {1\over 192}
\sig^{IJKL}_{ab} {\cal S}^a(z) {\cal S}^b(w)~.
\label{A-spinor}
\eea
Substituting these expressions back into eq.(\ref{lhs-show-OPE}) one
almost establishes eq.(\ref{show-OPE}) except for the fact that the
last equation in (\ref{A-spinor}) tells us how to write self-dual
tensors in terms of the spinors and not the ones which are anti-self-dual.
It would be interesting to understand this problem better through an
explicit bosonisation.

\sectiono{A standard regularisation}
\label{a:regularization}

Here we shall discuss a standard way to regularise the expression in
(\ref{h0-form}).
One may start with the following result derived in \cite{ambjorn81}:
\bea && \pi^{-s/2} \Gamma(s/2) \sum_{n=-\infty}^{\infty} \lt[
\lt(m\over \pi\rt)^2 + \lt(n \over a\rt)^2 \rt]^{-s/2} \cr &=&
{am^{1-s} \over \pi^{(1-s)/2}} \lt[ \Gamma(s/2 -1/2) + 2
\sum_{n=-\infty}^{\infty~\prime} {K_{1/2-s/2} (2m|an|) \over
(m|an|)^{1/2-s/2}} \rt]~, \label{analytics} \eea which is valid for
negative $s$. The prime on the summation in the second term on the
right hand side indicates that the $n=0$ contribution is excluded.
Setting $s=-1$ one can derive: \bea \sum_{n=1}^{\infty} \lt[
\lt(n\over a \rt)^2 + \lt(m\over 2\pi \rt)^2 \rt]^{1/2} = - {am^2
\over 16 \pi^2} \Gamma(-1) -{m\over 4\pi} - {m\over 2\pi^2}
\sum_{n=1}^{\infty} {1\over n} K_1(amn)~, \label{s=-1} \eea where
$K_{\nu}(x)$ is a modified Bessel function. The Casimir energy of a
one dimensional massive scalar field with mass $m$ can be obtained
from the above result where $a$ is proportional to the size of the
system. The first term on the right hand side is divergent and
corresponds to a constant energy density. Dropping it off is the
standard regularisation. The second term is independent of the size
of the system - ignoring this is a constant shift of the total
energy \cite{ambjorn81}. Therefore we may write: \bea
\sum_{n=1}^{\infty} \lt[ \lt(n\over a \rt)^2 + \lt(m\over 2\pi
\rt)^2 \rt]^{1/2}_{reg.} = - {m\over 2\pi^2} \sum_{n=1}^{\infty}
{1\over n} K_1(amn)~, \label{s=-1reg} \eea Now noticing that the
right hand side of eq.(\ref{h0-form}) can be written in the
following form, \bea h_0(m) = 8 \sum_{n=1}^{\infty}
\sqrt{n^2+\lt(m\over 2\pi \rt)^2} - 4 \sum_{n=1}^{\infty}
\sqrt{\lt(n\over 2\rt)^2 + \lt(m\over 2\pi
  \rt)^2}~,
\eea and using eq.(\ref{s=-1reg}) one gets the following result:
\bea [h_0(m)]_{reg.} = {2m\over \pi^2} \sum_{n=1}^{\infty} {1\over
  n} \lt[K_1(2mn)-2K_1(mn) \rt] ~.
\eea Notice that using the limiting behaviour: $\lim_{x\to 0}
xK_1(x) \to 1$ one gets the correct massless limit: \bea \lim_{m\to
0} [h_0(m)]_{reg.} \to -{1\over 2}~. \eea


\begin{thebibliography}{99}

\bibitem{sen04}
A.~Sen,
``Tachyon dynamics in open string theory,''
Int.\ J.\ Mod.\ Phys.\ A {\bf 20}, 5513 (2005)
[arXiv:hep-th/0410103].

\bibitem{lit}
N.~Drukker, D.~J.~Gross and N.~Itzhaki,
``Sphalerons, merons and unstable branes in AdS,''
Phys.\ Rev.\ D {\bf 62}, 086007 (2000)
[arXiv:hep-th/0004131];
K.~Peeters and M.~Zamaklar,
``AdS/CFT description of D-particle decay,''
Phys.\ Rev.\ D {\bf 71}, 026007 (2005)
[arXiv:hep-th/0405125];
Y.~L.~He and P.~Zhang,
``ZZ brane decay in D dimensions,''
JHEP {\bf 0611}, 014 (2006)
[arXiv:hep-th/0607188];
D.~Israel and E.~Rabinovici,
``Rolling tachyon in anti-de Sitter space-time,''
arXiv:hep-th/0609087.

\bibitem{universal}
A.~Sen,
``Universality of the tachyon potential,''
JHEP {\bf 9912}, 027 (1999)
[arXiv:hep-th/9911116].

\bibitem{ads/cft}
J.~M.~Maldacena,
``The large N limit of superconformal field theories and supergravity,''
Adv.\ Theor.\ Math.\ Phys.\  {\bf 2}, 231 (1998)
[Int.\ J.\ Theor.\ Phys.\  {\bf 38}, 1113 (1999)]
[arXiv:hep-th/9711200];
S.~S.~Gubser, I.~R.~Klebanov and A.~M.~Polyakov,
``Gauge theory correlators from non-critical string theory,''
Phys.\ Lett.\ B {\bf 428}, 105 (1998)
[arXiv:hep-th/9802109];
E.~Witten,
``Anti-de Sitter space and holography,''
Adv.\ Theor.\ Math.\ Phys.\  {\bf 2}, 253 (1998)
[arXiv:hep-th/9802150].

\bibitem{blau01}
M.~Blau, J.~Figueroa-O'Farrill, C.~Hull and G.~Papadopoulos,
``A new maximally supersymmetric background of IIB superstring theory,''
JHEP {\bf 0201}, 047 (2002)
[arXiv:hep-th/0110242].

\bibitem{metsaev01}
R.~R.~Metsaev,
``Type IIB Green-Schwarz superstring in plane wave Ramond-Ramond
background,''
Nucl.\ Phys.\ B {\bf 625}, 70 (2002)
[arXiv:hep-th/0112044].

\bibitem{metsaev02}
R.~R.~Metsaev and A.~A.~Tseytlin,
``Exactly solvable model of superstring in plane wave Ramond-Ramond
background,''
Phys.\ Rev.\ D {\bf 65}, 126004 (2002)
[arXiv:hep-th/0202109].

\bibitem{bmn}
D.~Berenstein, J.~M.~Maldacena and H.~Nastase,
``Strings in flat space and pp waves from N = 4 super Yang Mills,''
JHEP {\bf 0204}, 013 (2002)
[arXiv:hep-th/0202021].

\bibitem{gsw}
M.~B.~Green, J.~H.~Schwarz and E.~Witten,
``SUPERSTRING THEORY. VOL. 1: INTRODUCTION,''

\bibitem{bps}
J.~Dai, R.~G.~Leigh and J.~Polchinski,
``NEW CONNECTIONS BETWEEN STRING THEORIES,''
Mod.\ Phys.\ Lett.\ A {\bf 4}, 2073 (1989);
J.~Polchinski, S.~Chaudhuri and C.~V.~Johnson,
``Notes on D-Branes,''
arXiv:hep-th/9602052;
J.~Polchinski,
``Lectures on D-branes,''
arXiv:hep-th/9611050;

\bibitem{nbps}
A.~Sen,
``Non-BPS states and branes in string theory,''
arXiv:hep-th/9904207;
A.~Lerda and R.~Russo,
``Stable non-BPS states in string theory: A pedagogical review,''
Int.\ J.\ Mod.\ Phys.\ A {\bf 15}, 771 (2000)
[arXiv:hep-th/9905006];
O.~Bergman and M.~R.~Gaberdiel,
``Non-BPS Dirichlet branes,''
Class.\ Quant.\ Grav.\  {\bf 17}, 961 (2000)
[arXiv:hep-th/9908126];
J.~H.~Schwarz,
``TASI lectures on non-BPS D-brane systems,''
arXiv:hep-th/9908144;
M.~R.~Gaberdiel,
``Lectures on non-BPS Dirichlet branes,''
Class.\ Quant.\ Grav.\  {\bf 17}, 3483 (2000)
[arXiv:hep-th/0005029].

\bibitem{dabholkar02}
A.~Dabholkar and S.~Parvizi,
``Dp branes in pp-wave background,''
Nucl.\ Phys.\ B {\bf 641}, 223 (2002)
[arXiv:hep-th/0203231].

\bibitem{bergman02}
O.~Bergman, M.~R.~Gaberdiel and M.~B.~Green,
``D-brane interactions in type IIB plane-wave background,''
JHEP {\bf 0303}, 002 (2003)
[arXiv:hep-th/0205183];
M.~R.~Gaberdiel and M.~B.~Green,
``The D-instanton and other supersymmetric D-branes in IIB plane-wave string
Annals Phys.\  {\bf 307}, 147 (2003)
[arXiv:hep-th/0211122];
M.~R.~Gaberdiel, M.~B.~Green, S.~Schafer-Nameki and A.~Sinha,
``Oblique and curved D-branes in IIB plane-wave string theory,''
JHEP {\bf 0310}, 052 (2003)
[arXiv:hep-th/0306056].

\bibitem{skenderis02}
K.~Skenderis and M.~Taylor,
``Open strings in the plane wave background. I: Quantization and
symmetries,''
Nucl.\ Phys.\ B {\bf 665}, 3 (2003)
[arXiv:hep-th/0211011];
K.~Skenderis and M.~Taylor,
``Open strings in the plane wave background. II: Superalgebras and spectra,''
JHEP {\bf 0307}, 006 (2003)
[arXiv:hep-th/0212184];
K.~Skenderis and M.~Taylor,
``Properties of branes in curved spacetimes,''
JHEP {\bf 0402}, 030 (2004)
[arXiv:hep-th/0311079].

\bibitem{ws}
M.~Billo and I.~Pesando,
``Boundary states for GS superstrings in an Hpp wave background,''
Phys.\ Lett.\ B {\bf 536}, 121 (2002)
[arXiv:hep-th/0203028];
P.~Bain, K.~Peeters and M.~Zamaklar,
``D-branes in a plane wave from covariant open strings,''
Phys.\ Rev.\ D {\bf 67}, 066001 (2003)
[arXiv:hep-th/0208038];
Y.~Hikida and S.~Yamaguchi,
``D-branes in pp-waves and massive theories on worldsheet with boundary,''
JHEP {\bf 0301}, 072 (2003)
[arXiv:hep-th/0210262];
J.~w.~Kim, B.~H.~Lee and H.~S.~Yang,
``Superstrings and D-branes in plane wave,''
Phys.\ Rev.\ D {\bf 68}, 026004 (2003)
[arXiv:hep-th/0302060];
K.~S.~Cha, B.~H.~Lee and H.~S.~Yang,
``Intersecting D-branes in type IIB plane wave background,''
Phys.\ Rev.\ D {\bf 68}, 106004 (2003)
[arXiv:hep-th/0307146];
C.~Y.~Park,
``Open string spectrum in pp-wave background,''
J.\ Korean Phys.\ Soc.\  {\bf 44}, 235 (2004)
[arXiv:hep-th/0308151];
K.~S.~Cha, B.~H.~Lee and H.~S.~Yang,
``A complete classification of D-branes in type IIB plane wave  background,''
JHEP {\bf 0403}, 058 (2004)
[arXiv:hep-th/0310177];
M.~Sakaguchi and K.~Yoshida,
``D-branes of covariant AdS superstrings,''
Nucl.\ Phys.\ B {\bf 684}, 100 (2004)
[arXiv:hep-th/0310228];
T.~Mattik,
``Branes in the plane wave background with gauge field condensates,''
JHEP {\bf 0506}, 041 (2005)
[arXiv:hep-th/0501088];
B.~H.~Lee, J.~w.~Lee, C.~Park and H.~S.~Yang,
``More on supersymmetric D-branes in type IIB plane wave background,''
JHEP {\bf 0601}, 015 (2006)
[arXiv:hep-th/0506091];
T.~Mattik,
``Boundary fermions and the plane wave,''
JHEP {\bf 0605}, 045 (2006)
[arXiv:hep-th/0603153].

\bibitem{le}
K.~Skenderis and M.~Taylor,
``Branes in AdS and pp-wave spacetimes,''
JHEP {\bf 0206}, 025 (2002)
[arXiv:hep-th/0204054];
P.~Bain, P.~Meessen and M.~Zamaklar,
``Supergravity solutions for D-branes in Hpp-wave backgrounds,''
Class.\ Quant.\ Grav.\  {\bf 20}, 913 (2003)
[arXiv:hep-th/0205106];
S.~S.~Pal,
``Solution to worldvolume action of D3 brane in pp-wave background,''
Mod.\ Phys.\ Lett.\ A {\bf 17}, 1735 (2002)
[arXiv:hep-th/0205303];
R.~R.~Metsaev,
``Supersymmetric D3 brane and N = 4 SYM actions in plane wave backgrounds,''
Nucl.\ Phys.\ B {\bf 655}, 3 (2003)
[arXiv:hep-th/0211178].

\bibitem{green96}
M.~B.~Green and M.~Gutperle,
``Light-cone supersymmetry and D-branes,''
Nucl.\ Phys.\ B {\bf 476}, 484 (1996)
[arXiv:hep-th/9604091].

\bibitem{lambert99}
N.~D.~Lambert and P.~C.~West,
``D-branes in the Green-Schwarz formalism,''
Phys.\ Lett.\ B {\bf 459}, 515 (1999)
[arXiv:hep-th/9905031].

\bibitem{nbps1}
P.~Mukhopadhyay,
``Unstable non-BPS D-branes of type-II string theories in light-cone
Green-Schwarz formalism,''
Nucl.\ Phys.\ B {\bf 600}, 285 (2001)
[arXiv:hep-th/0011047].

\bibitem{bs}
N. V. Suryanarayana, August 2004 - unpublished.

\bibitem{nbps2}
P.~Mukhopadhyay,
``Non-Bps D-Branes In Light-Cone Green-Schwarz Formalism,''
JHEP {\bf 0501}, 059 (2005)
[arXiv:hep-th/0411103].

\bibitem{yoneya99}
T.~Yoneya,
``Spontaneously broken space-time supersymmetry in open string theory
without GSO projection,''
Nucl.\ Phys.\ B {\bf 576}, 219 (2000)
[arXiv:hep-th/9912255].

\bibitem{berkovits}
N.~Berkovits,
``Super-Poincare covariant quantization of the superstring,''
JHEP {\bf 0004}, 018 (2000)
[arXiv:hep-th/0001035].

\bibitem{ps1}
P.~Mukhopadhyay,
``On D-Brane Boundary State Analysis In Pure-Spinor Formalism,''
JHEP {\bf 0603}, 066 (2006)
[arXiv:hep-th/0505157].

\bibitem{ps2}
P.~Mukhopadhyay,
``DDF construction and D-brane boundary states in pure spinor formalism,''
JHEP {\bf 0605}, 055 (2006)
[arXiv:hep-th/0512161].

\bibitem{nBPS-RR}
P.~Mukhopadhyay,
``Tachyon condensation and "non-BPS" D-branes in a
Ramond-Ramond plane wave background,''
Talk presented at Great Lakes Strings Conference, March-April, 2006.

\bibitem{us}
P.~Mukhopadhyay, ``A Universality in PP-Waves,'' JHEP {\bf 0706},
061 (2007) [arXiv:0704.0085 [hep-th]].

\bibitem{action}
A.~Sen,
``Supersymmetric world-volume action for non-BPS D-branes,''
JHEP {\bf 9910}, 008 (1999)
[arXiv:hep-th/9909062].
M.~R.~Garousi,
``Tachyon couplings on non-BPS D-branes and Dirac-Born-Infeld action,''
Nucl.\ Phys.\ B {\bf 584}, 284 (2000)
[arXiv:hep-th/0003122].
E.~A.~Bergshoeff, M.~de Roo, T.~C.~de Wit, E.~Eyras and S.~Panda,
``T-duality and actions for non-BPS D-branes,''
JHEP {\bf 0005}, 009 (2000)
[arXiv:hep-th/0003221];
J.~Kluson,
``Proposal for non-BPS D-brane action,''
Phys.\ Rev.\ D {\bf 62}, 126003 (2000)
[arXiv:hep-th/0004106].


\bibitem{RR}
E.~Witten,
``D-branes and K-theory,''
JHEP {\bf 9812}, 019 (1998)
[arXiv:hep-th/9810188];
A.~Sen,
``BPS D-branes on non-supersymmetric cycles,''
JHEP {\bf 9812}, 021 (1998)
[arXiv:hep-th/9812031];
P.~Horava,
``Type IIA D-branes, K-theory, and matrix theory,''
Adv.\ Theor.\ Math.\ Phys.\  {\bf 2}, 1373 (1999)
[arXiv:hep-th/9812135];
A.~Sen,
``Non-BPS states and branes in string theory,''
arXiv:hep-th/9904207.

\bibitem{lc}
G.~T.~Horowitz and A.~R.~Steif,
``STRINGS IN STRONG GRAVITATIONAL FIELDS,''
Phys.\ Rev.\ D {\bf 42}, 1950 (1990).
R.~E.~Rudd,
``Light cone gauge quantization of 2-D sigma models,''
Nucl.\ Phys.\ B {\bf 427}, 81 (1994)
[arXiv:hep-th/9402106].

\bibitem{maldacena02}
J.~M.~Maldacena and L.~Maoz,
``Strings on pp-waves and massive two dimensional field theories,''
JHEP {\bf 0212}, 046 (2002)
[arXiv:hep-th/0207284];


\bibitem{generalization}
J.~G.~Russo and A.~A.~Tseytlin,
``Constant magnetic field in closed string theory: An Exactly
solvable model,''
%
Nucl.\ Phys.\ B {\bf 448}, 293 (1995)
[arXiv:hep-th/9411099];
J.~G.~Russo and A.~A.~Tseytlin,
``On solvable models of type IIB superstring in NS-NS and R-R plane
wave backgrounds,''
%
JHEP {\bf 0204}, 021 (2002)
[arXiv:hep-th/0202179];
J.~G.~Russo and A.~A.~Tseytlin,
``A class of exact pp-wave string models with interacting light-cone
gauge actions,''
%
JHEP {\bf 0209}, 035 (2002)
[arXiv:hep-th/0208114];

\bibitem{papadopoulos02}
G.~Papadopoulos, J.~G.~Russo and A.~A.~Tseytlin,
``Solvable model of strings in a time-dependent plane-wave
background,''
%
Class.\ Quant.\ Grav.\  {\bf 20}, 969 (2003)
[arXiv:hep-th/0211289].


\bibitem{friedan85}
D.~Friedan, E.~J.~Martinec and S.~H.~Shenker,
``Conformal Invariance, Supersymmetry And String Theory,''
Nucl.\ Phys.\ B {\bf 271}, 93 (1986).

\bibitem{kostelecky86}
V.~A.~Kostelecky, O.~Lechtenfeld, W.~Lerche, S.~Samuel and S.~Watamura,
``CONFORMAL TECHNIQUES, BOSONIZATION AND TREE LEVEL STRING AMPLITUDES,''
Nucl.\ Phys.\ B {\bf 288}, 173 (1987).

\bibitem{witten84}
E.~Witten,
``D = 10 Superstring Theory,''


\bibitem{ambjorn81}
J.~Ambjorn and S.~Wolfram,
``Properties Of The Vacuum. 1. Mechanical And Thermodynamic,''
Annals Phys.\  {\bf 147}, 1 (1983).


\end{thebibliography}
\end{document}